\documentclass[usenatbib]{mn2e} 

\usepackage{amssymb} \usepackage{epsfig}
\usepackage{graphicx}

\def\lsim{\mathrel{\rlap{\lower4pt\hbox{\hskip1pt$\sim$}}
    \raise1pt\hbox{$<$}}}                
\def\gsim{\mathrel{\rlap{\lower4pt\hbox{\hskip1pt$\sim$}}
    \raise1pt\hbox{$>$}}}                

\begin{document}

\title[Cosmic Reionisation and the Ly$\alpha$ Emitting Fraction at
  High Redshift] {Keck Spectroscopy of Faint $3<z<7$ Lyman Break
  Galaxies: - I. New constraints on cosmic reionisation from the
  luminosity and redshift-dependent fraction of Lyman-$\alpha$
  emission }

\author[Stark et al.]  {Daniel
  P. Stark$^{1}$\thanks{dps@ast.cam.ac.uk}, Richard S. Ellis$^{2}$,
  Kuenley Chiu$^{2}$, Masami Ouchi$^{3,4}$, Andrew Bunker$^{5}$
  \\ $^{1}$ Kavli Institute of Cosmology \&  Institute of Astronomy,
  University of Cambridge, Madingley Road, Cambridge CB3 0HA, UK
  \\   $^{2}$ California Institute of Technology, 1200 E. Calfornia
  Blvd. Pasadena, CA 91125, USA \\ $^{3}$ Observatories of the
  Carnegie Institution of Washington, 813 Santa Barbara Street
  Pasadena, CA 91101 USA \\ $^{4}$ Carnegie Fellow \\ $^{5}$
  Department of Physics, University of Oxford, Denys Wilkinson
  Building, Keble Road, Oxford OX1 3RH, UK} \date{Accepted ... ;
  Received ... ; in original form ...}

\pagerange{\pageref{firstpage}--\pageref{lastpage}} \pubyear{2010}

\hsize=6truein \maketitle
\label{firstpage}

\begin{abstract}

We present the first results of a new Keck spectroscopic survey of UV
faint Lyman break galaxies in the redshift range $3<z<7$. Combined
with earlier Keck and published ESO VLT data, our spectroscopic sample
contains more than 600 dropouts offering new insight into the  nature
of sub-$L^{\ast}$ sources typical of those likely to dominate the
cosmic reionisation process.  In this first paper in a series
discussing these observations, we characterise the fraction of strong
Ly$\alpha$ emitters within the continuum-selected dropout population.
By quantifying how the ``Ly$\alpha$ fraction'' varies with redshift,
we seek to constrain changes in Ly$\alpha$ transmission associated
with reionisation.  In order to distinguish the effects  of
reionisation from other factors which affect the Ly$\alpha$ fraction
(e.g. dust, ISM kinematics), we study the luminosity and
redshift-dependence of the Ly$\alpha$ fraction  over $3\lsim z\lsim
6$, when the IGM is known to be ionised.  These  results reveal that
low luminosity galaxies show strong Ly$\alpha$ emission much more
frequently than luminous systems, and that at fixed  luminosity, the
prevalence of strong Lyman-$\alpha$ emission increases  moderately
with redshift over $3 < z <6$.  Based on  the striking correlation
between blue UV slopes and strong Ly$\alpha$  emitting galaxies in our
dataset, we argue that the Ly$\alpha$ fraction trends are  governed by
redshift and luminosity-dependent  variations in the dust obscuration,
with likely additional contributions from trends in the  kinematics
and covering fraction of neutral hydrogen.  We find a tentative
decrease in the  Ly$\alpha$ fraction at $z\simeq 7$ based on the
limited IR spectroscopic data for candidate $z\simeq 7$ galaxies, a
result which, if confirmed with future surveys, would suggest an
increase in the neutral fraction by this epoch. Given the abundant
supply of $z$ and Y-drops now available from deep Hubble WFC3/IR
surveys, we show it will soon be possible to significantly improve
estimates of the Ly$\alpha$ fraction using optical and near-infrared
multi-object spectrographs, thereby extending the study conducted in
this paper to $7\lsim z\lsim 8$.

\end{abstract}


\section{Introduction}

Considerable observational progress has been achieved in recent years
in the study of star-forming galaxies seen beyond $z\simeq 3$, a
period corresponding  to $\simeq 2$ Gyr after the Big Bang. It is now
clear that this is a period of rapid galaxy evolution and a number of
key results have emerged from recent multi-wavelength surveys.

For the colour-selected $z>3$ {\it Lyman break galaxies} (LBGs), it is
now established from various independent surveys that the star
formation density, deduced from rest-frame UV luminosities, declines
monotonically with redshift (e.g. \citealt{Stanway03,Bunker04})
largely as a result of a corresponding fading of the characteristic UV
luminosity (e.g., \citealt{Ouchi04a,
  Yoshida06,Bouwens06a,Bouwens07,McLure10}) The associated stellar
mass density in Lyman break galaxies,  deduced from near-infrared
Spitzer photometry, increases by $\simeq$1 dex from $z\simeq 6$ to 4
\citep{Eyles07,Stark09}. As  the rate of change of stellar mass is
governed by ongoing star formation, it is useful to relate the two
measures and such a comparison indicates a rapid duty-cycle of star
formation activity at this time, unlike the more continuous modes seen
for equivalent sources at $z\simeq 2-3$ \citep{Stark09}. By contrast,
the redshift-dependent luminosity function of  narrow-band selected
{\it Lyman alpha emitters}, shows no equivalent decline  with redshift
over $3 < z < 6$ \citep{Ouchi08}, suggesting an increasing fraction of
line emitters amongst the star forming population at early
times. Moreover, detailed  studies of the slope of the UV continuum in
$z>3$ Lyman break galaxies indicates  a decreasing dust content at
earlier times \citep[e.g.][]{Stanway05,Bouwens06a,Bouwens09b} as well
as a luminosity dependence at $z\simeq 3$ \citep[e.g.][]{Reddy09}.
Conceivably the combination of a reduced dust content and a shift to
more intense, shorter-term star formation at high redshift, can
explain these various redshift-dependent trends.   

Notwithstanding this considerable progress, a major concern is that
the above conclusions  rest largely on deductions made with {\it
  photometric data}, particularly for the Lyman break population.
Quite apart from the possibility of low redshift interlopers lying
within the photometric samples (a problem that increases for drop-out
selected samples at redder wavelengths), the wanted physical measures
of the star formation rate,  dust content and stellar mass are all
rendered uncertain by the absence of  precise spectroscopic
redshifts. While considerable effort has been invested in the
spectroscopic study of $z\simeq 3$ Lyman break galaxies
\citep[e.g.][]{Shapley03,Steidel03,Quider10}, comparably little
spectroscopy has been achieved for higher redshift samples.
\citet{Steidel99} obtained spectroscopic redshifts for nearly  50
bright ($I<25$) $z\simeq 4$ LBGs.  Most surveys of $z\simeq 5-6$ $V$
and $i'$-drops have generally involved relatively small samples,
typically comprising fewer than 10 sources
\citep[e.g.][]{Stanway04,Stanway07,Ando07,Dow-Hygelund07}.  Recent
deep HST {\it ACS} Grism  observations of the Hubble Ultra Deep Field
have allowed the spectra  of faint $z\simeq 5$ LBGs to be
characterised \citep{Rhoads09}, albeit at very low spectral
resolution, resulting in 39 redshift confirmations.  The largest
spectroscopic sample of $4<z<6$ LBGs thus far published is the
VLT/FORS2 survey of the Chandra Deep Field South
\citep{Vanzella02,Vanzella05,Vanzella06,Vanzella08,Vanzella09}.  This
survey represents a major step forward, targetting 195 $B$, $V$ and
$i'$-drop galaxies and securing high redshifts for 99 of them.  In
addition,  a recent campaign with the VIMOS spectrograph on the VLT
has also targeted  bright $z\gsim 3.5$ LBGs ($i'_{775}<\rm{25}$) in
the Chandra Deep Field  South, confirming redshifts for 20 bright
sources at $3.5<z<5.5$ \citep{Balestra10}. 

We build on these strides in this paper, the first in a series
presenting the results of a new Keck survey of $3<z<7$ LBGs selected
photometrically in the GOODS fields. The overall goal is to improve
our understanding of the evolution of star-forming galaxies during the
first 2 Gyr of cosmic history. Fully exploiting the 10 metre Keck
aperture, we have designed our survey to target  intrinsically fainter
sources than those reached in the VLT/FORS2 survey, thereby
complementing that effort. Spectroscopy spanning a wide range of
intrinsic luminosities is very important if we seek to understand
earlier examples of the luminosity-dependent trends seen at $z\simeq
3$ \citep{Reddy09}.  As discussed below, it is equally important to
target and study sub-luminous  star forming sources at early times, as
these may be typical of those  galaxies responsible for cosmic
reionisation \citep{Bouwens07,Ouchi09,Bunker10,Oesch10,McLure10}.

Understanding evolution in the demographic trends of star-forming
galaxies  over $3<z<7$ is vital if we are to use galaxies as tracers
of {\it cosmic reionisation}, a cosmic event which is now a frontier
of observational cosmology.  Knowledge of when reionisation occurred
is crucial to our understanding  of the nature of the earliest
UV-emitting sources as well as the discrepancy between the observed
number of dwarf galaxies and those expected from cosmological
simulations \citep[e.g.][]{Salvadori09}.  Current observational
constraints are limited.  WMAP measurements indicate that the universe
may  have been partially ionised as early as $z\simeq 11$
\citep{Dunkley09, Larson10},  while observations of transmission in
the spectra of quasars  reveal that intergalactic hydrogen must be
highly ionised below $z\lsim 6$.  While the  discovery of an
accelerated decline and increased variance in the mean  transmitted
flux from quasars initially led many to suggest that the intergalactic
medium (IGM) is still partially neutral at  $z\simeq 6.2$
\citep{Fan06}, recent work has demonstated that these results do not
necessarily require a sudden change in  the ionisation state of the
IGM \citep{Becker07}. Given the difficulty in locating quasars at
$z>7$,  it seems unlikely that quasar spectroscopy will constrain the
epoch  when the bulk of the IGM was reionised in the near future. 

Ly$\alpha$ emitting galaxies offer a valuable additional probe of the
IGM (e.g., \citealt{Rhoads01,Malhotra04,Kashikawa06}). In principle,
the test is straightforward to apply.  Young galaxies  emit  copious
amounts of Ly$\alpha$ photons, which are resonantly scattered by
neutral hydrogen.  Hence as we probe the regime when the IGM becomes
significantly neutral, the fraction of star-forming galaxies showing
strong Ly$\alpha$ emission should decrease
\citep[e.g.][]{Haiman99,Santos04a,Furlanetto06,McQuinn07,Mesinger08,Iliev08,
  Dayal10}.  Recent  measurements of the luminosity function of
Ly$\alpha$ emitters (LAEs) selected via narrowband imaging have
revealed a tantalising decline between $z=5.7$ and  $z=7.0$
\citep[e.g.][]{Kashikawa06,Iye06,Ota08}, offering possible evidence
that  the ionisation state of the IGM evolves over $6<z<7$.  

Regardless of the validity of claims for an increasing fraction of
neutral hydrogen over $6<z<7$, in practice the interpretation of the
Ly$\alpha$ test  is more complex.  While the ionisation state of the
IGM  affects the Ly$\alpha$ LF, so does evolution of a multitude of
other properties intrinsic to the sampled population
(e.g., \citealt{Verhamme06, Verhamme08, Atek08}).  Evolution in the dust content
\citep{Bouwens09b}, the column density, kinematics, and geometrical
distribution  (generally described as the ``covering fraction'') of
neutral hydrogen  \citep{Shapley03,Quider09,Quider10, Steidel10} can
each play a key role.  Also important for the transmission of
Ly$\alpha$  photons is the density of the IGM,  which evolves
continuously with redshift, and the stellar initial mass function  of
galaxies, for which few robust constraints  exist at high-redshift.  

The existence of these complicating factors highlights the importance
of understanding how the prevalence of Ly$\alpha$-emitting galaxies
varies just after reionisation in addition to characterising the
decline that may occur during reionisation itself.  With this goal in
mind, we seek to construct an independent measure of the redshift
evolution of  Ly$\alpha$ emitting galaxies over a large redshift
baseline, complementary to past efforts \citep[e.g.][]{Ota08}. Here we
introduce the method of measuring the fraction of strong Ly$\alpha$
emitters (hereafter the ``Ly$\alpha$ fraction'') within the LBG
population.  Applying the Ly$\alpha$ fraction test to a large sample
of LBGs  has many advantages.  Firstly, the LBG samples are already in
place over $4<z<8$ owing to deep surveys with HST \citep[e.g.][]
{McLure10,Bunker10,Bouwens10a}; hence the only time investment
required is follow-up spectroscopy. This spectroscopy not only
provides  a sample of Ly$\alpha$ emitters with known spectroscopic
redshifts, but it also provides  information on the kinematics of the
ISM \citep[e.g.][]{Shapley03,Vanzella09,Steidel10} for   individual
bright sources (and for faint systems via composite spectra)  and
improves estimates of dust obscuration via UV colors, both of which
are  key factors governing the transmission of Ly$\alpha$ photons.  By
improving our understanding of how these properties change with time,
we can begin to isolate the effect of reionization on the evolution in
the Ly$\alpha$ fraction.  Additionally, the Ly$\alpha$ fraction is
insensitive to the declining number density of star-forming galaxies,
in  contrast to the Ly$\alpha$ luminosity function test (which
requires  comparison to the UV luminosity function to account for this
degeneracy).  Naturally, the Ly$\alpha$ fraction test has its own
complications, but we show that these can be corrected for (see \S3.3
and 3.4), and we thus argue that this test will provide valuable
constraints on reionization as new $7<z<10$ LBGs emerge in the next
several years.

Our goal is therefore to obtain a robust measure of the luminosity and
redshift dependence of the Ly$\alpha$ fraction when the IGM is
highly-ionised ($z\simeq 3-6$) and to compare it to the Ly$\alpha$
fraction at progressively earlier times.  If the IGM ionisation state
evolves significantly at $z\gsim 6$ as possibly implied by the
narrowband Ly$\alpha$ results, we would expect the measured fraction
of LBGs with strong Ly$\alpha$ emission to be lower at $z\simeq 6-8$
than expected from extrapolating the trends seen over $3<z<6$.  By
placing the evolution of the Ly$\alpha$ fraction at $3<z<6$ in the
context of the evolution of the well-characterised  LBG parent
population (e.g. variation of dust extinction  with redshift and
luminosity), we will calibrate the relative importance of factors
other than reionisation on the transmission of Ly$\alpha$ photons.

Our Ly$\alpha$ fraction test is ideally suited to our  large
spectroscopic sample of continuum-selected LBGs spanning  the redshift
range $3\lsim z\lsim7$. The current Keck sample (including a sample of
archival $i'$-drops  to be presented in Bunker et al. 2010, in
preparation) consists of 455 B, V, $i'$, and $z$-drops
(photometrically-selected to lie at $3.5<z<7.0$) spanning a wide range
in UV luminosity (to M$_{\rm{UV}}\simeq -18$).  We combine this sample
with a more luminous publically-available FORS2 dataset of 195 sources
satisfying photometric criteria similar to those adopted for the Keck
surveys.  Using this large spectroscopic database, we address the
primary goal of  this first paper in our series - to compute the
Ly$\alpha$ fraction  as a function of luminosity and redshift. This
will  enable us to identify the principal factors governing its
evolution over $3\lsim z\lsim 6$, prior to extending the test to
$6\lsim z\lsim 8$, where we can probe changes in the IGM.  

The plan of the paper is as follows.  In \S2, we describe the target
selection and spectroscopic observations, and present the rest-frame
UV properties of our current sample.  In \S3, we describe the
construction  of the Ly$\alpha$ catalog, discuss the method used to
measure Ly$\alpha$  equivalent widths, and compute the completeness of
Ly$\alpha$ detection  as a function of luminosity and redshift.  In
\S4,  we discuss the luminosity dependence  of Ly$\alpha$ emission in
the context of earlier studies at $z\simeq 3$. We then turn to the
redshift-dependence of Ly$\alpha$ emission and discuss the important
implications of our findings. Finally, in \S5, we examine the extant
data on the rate of occurrence of Ly$\alpha$ emission for candidate
sources thought to lie beyond $z\simeq 6$. We use this to test the
practicality of using our test as a valuable probe of cosmic
reionisation. We summarise the conclusions of our study in \S6.

Throughout the paper, we adopt a $\Lambda$-dominated, flat universe
with $\Omega_{\Lambda}=0.7$, $\Omega_{M}=0.3$ and
$\rm{H_{0}}=70\,\rm{h_{70}}~{\rm km\,s}^{-1}\,{\rm Mpc}^{-1}$. All
magnitudes in this paper are quoted in the AB system \citep{Oke83}.

\section{Observations}

We present the results of a new and ongoing Keck spectroscopic survey
of photometrically-selected $B$, $V$ and $i'$-band `dropouts' in the
northern and southern GOODS fields \citep{Giavalisco04a}. The GOODS
fields were selected for this survey on account of the depth and
precision of their unique multi-color  photometric data useful for
selecting targets, as well as the availability of associated Spitzer
and Chandra data which provided valuable stellar masses and
AGN-related  properties
\citep{Eyles05,HYan06,Eyles07,Stark07a,Stark09,Gonzalez09,Labbe10}. 

\subsection{The Keck/DEIMOS Survey in GOODS-N and GOODS-S}

The majority of spectra discussed in this paper were obtained from an
ongoing  survey undertaken with the DEep Imaging Multi-Object
Spectrograph (DEIMOS) at the Nasmyth focus of the 10 m Keck II
telescope \citep{Faber03}.  DEIMOS is comprised of eight 2k x 4k CCDs
spanning roughly half of the ACS GOODS field of view ($\simeq 16\farcm
7 \times 5\farcm 0$) on the sky.  Our first observations have
primarily focused on targeting the $B$ and $V$-drop populations.
Although future DEIMOS  observations will expand coverage of higher
redshift $i'$ and $z$-band dropouts,  we include early data taken in
this territory, some of which is described independently in Bunker et al (2010,
in preparation).  

The recently-studied target list is primarily selected from the $B$,
$V$ and $i'$ dropout samples discussed in \citet{Stark09}.  To this we
will add earlier Keck data on $i'$ drops discussed in \S2.2
(\citealt{Bunker03,Stanway04}, Bunker et al. 2010, in preparation) as
well as newly-discovered $z$-drops from the recent WFC3/IR UDF
campaign (see below). Optical magnitudes range from relatively bright
systems ($z_{850}\simeq 23.5$) to the faintest   dropouts observed in
GOODS ($z_{850}\simeq 27.5$). Although the  ACS photometry in
\citet{Stark09} was based upon the GOODS version r1.1 ACS multi-band
source catalogs,  for this analysis we have updated the photometry to
the recently released GOODS version 2.0 catalogs.  These new catalogs
contain significantly deeper $z_{850}$-band imaging.  The 5$\sigma$
limiting magnitudes vary across the different  frames, but are
typically 28.0 in F435W (B$_{435}$), 28.2 in F606W (V$_{606}$),  27.9
in F775W ($i'_{775}$), and 27.8 in F850LP ($z_{850}$) when measured in
0.5 arcsecond diameter apertures.   At the median redshifts expected
for these  populations ($z=3.8$ for B-drops, $z=5.0$ for V-drops, and
$z=5.9$ for  $i'$-drops, e.g., Bouwens et al. 2007), these limits
correspond to absolute magnitude limits of M$_{\rm{UV}}\simeq -18.2$,
-18.8 , and -19.1 for B, V, and $i'$-drops.  

A summary of the various observing campaigns is given in Table 1.
Apart from the $i'$-drop exploratory survey discussed by Bunker et al.
2010, (in preparation),  in our campaigns during 2004-2007 targets
were selected to fill empty regions on slitmasks designed for other
purposes, e.g., studying the kinematics of disk galaxies at $z\simeq
1$ \citep{MacArthur08}.   For these  observing runs, we used the Gold
1200 line mm$^{-1}$ grating which provided coverage  between 5570 and
8210~\AA~ (allowing Ly$\alpha$ to be detected  in the redshift range
$z=3.6$ to 5.8), with a spectral pixel size of 0.3~\AA~ pixel$^{-1}$.
The spectral resolution measured from skylines was 1.4~\AA.  

\begin{table*}
\begin{tabular}{clcllccccll}
\hline  Number & Field & Mask ID & Date & $\rm{t_{exp}}$ (ksec) &
N$_{\rm{B}}$ &  N$_{\rm{V}}$ &   N$_{i}$ &  N$_{Z}$ & Grating
\\  \hline 

1 & GOODS-S & GS031 & 08-09 January 2003         &    19.8 & 0  & 0  &
3   &  0   & 1200 \\     2 & GOODS-N & GN031 & 02-06 April 2003 & 37.8
& 0  & 0  & 5  &  0   & 1200 \\     3 & GOODS-S & GS041 & 11 December
2004 & 21.9 & 0  &  0 & 20  &  0   & 1200 \\    4 & GOODS-S & GS051 &
31 October, 01-02 November 2005    & 8.4 & 0  & 0  & 17  &  0 & 1200
\\    5 & GOODS-S & GS071  & 10-11 November 2007 &    18.0 & 7 &  7 &
0  &  0  & 1200 \\     6 & GOODS-N & GN081  & 03-05 April 2008 & 21.6
& 85 &  9 & 0 &  0 & 600 \\     7 & GOODS-N & GN082  & 03-05 April
2008    & 21.6 & 95 & 12 & 0 &  0  & 600   \\     8 & GOODS-N & GN083
& 03-05 April 2008    & 20.4 & 86 & 14 & 0  &  0  & 600 \\     9 &
GOODS-N & GN094  & 24-26 March 2009    & 18.0 & 45 & 63 & 0 &  0  &
600  \\     10 & GOODS-N & GN095  & 24-26 March 2009 & 25.2 & 43 & 36
& 0  &  0  & 600  \\     11 & GOODS-S & GS091  & 19 October 2009 &
19.2 & 0  & 0 & 19 & 17 & 830  \\ \multicolumn{5}{l}{\hrulefill Total
  Unique Keck/DEIMOS Sample\hrulefill}  & 268  & 95 & 64 & 17 & $-$ &
\\ \multicolumn{5}{l}{\hrulefill Total Unique VLT/FORS2 Sample
  \hrulefill} & 83 & 56 & 56 & 0 & $-$ &
\\ \multicolumn{5}{l}{\hrulefill  Combined Unique Sample \hrulefill} &
351 & 151 & 108 & 17 & \\

\hline
\end{tabular}
\caption{Summary of observations with Keck/DEIMOS and
  VLT/FORS2. N$_{\rm{B}}$, N$_{\rm{V}}$,  N$_{\rm{i}}$, N$_{\rm{z}}$
  denote the number of B, V, $i'$, and $z$-drops  observed on each
  mask.  To maximise S/N, many sources were observed on multiple
  DEIMOS masks; we  account for such duplication when computing the
  total Keck sample size.  The $i'$-drops in the first two rows were
  originally published in Bunker et al. (2003) and Stanway et
  al. (2004), and the $i'$-drops presented in rows 3 and 4 are from
  Bunker et al. (2010, in preparation). The VLT FORS2 sample is take
  from observations presented in  Vanzella et al. (2009).  The
  combined total  sample in the last row includes all unique dropouts
  in the Keck and  VLT surveys.}
\end{table*}

The main survey began in earnest in 2008, when we started a dedicated
programme geared at obtaining spectra of $z>3$ dropouts.  Most spectra
were obtained  from 5 slitmasks observed in April 2008 and March 2009
(Table 1).  These slitmasks  utilised the 600 line mm$^{-1}$ grating
on DEIMOS, providing spectroscopic coverage between 4850~\AA and
10150~\AA~ (allowing Ly$\alpha$ to be detected between $z\simeq 3.0$
and 7.3) with a spectral pixel scale of 0.7 \AA~pixel$^{-1}$.  These
spectra provided a resolution (measured from skylines)  of $\simeq
3.5$~\AA.  This setup  allowed us to efficently follow-up B and V-band
dropouts simultaneously.   Each individual mask was observed for
$\simeq 5-7$ hours and contained  $\simeq 80$-110 dropouts.  Seeing
was typically 0\farcs 8, ranging  between 0\farcs 5 and 1\farcs 0.

\begin{figure*}
\begin{center}
\includegraphics[width=0.94\textwidth]{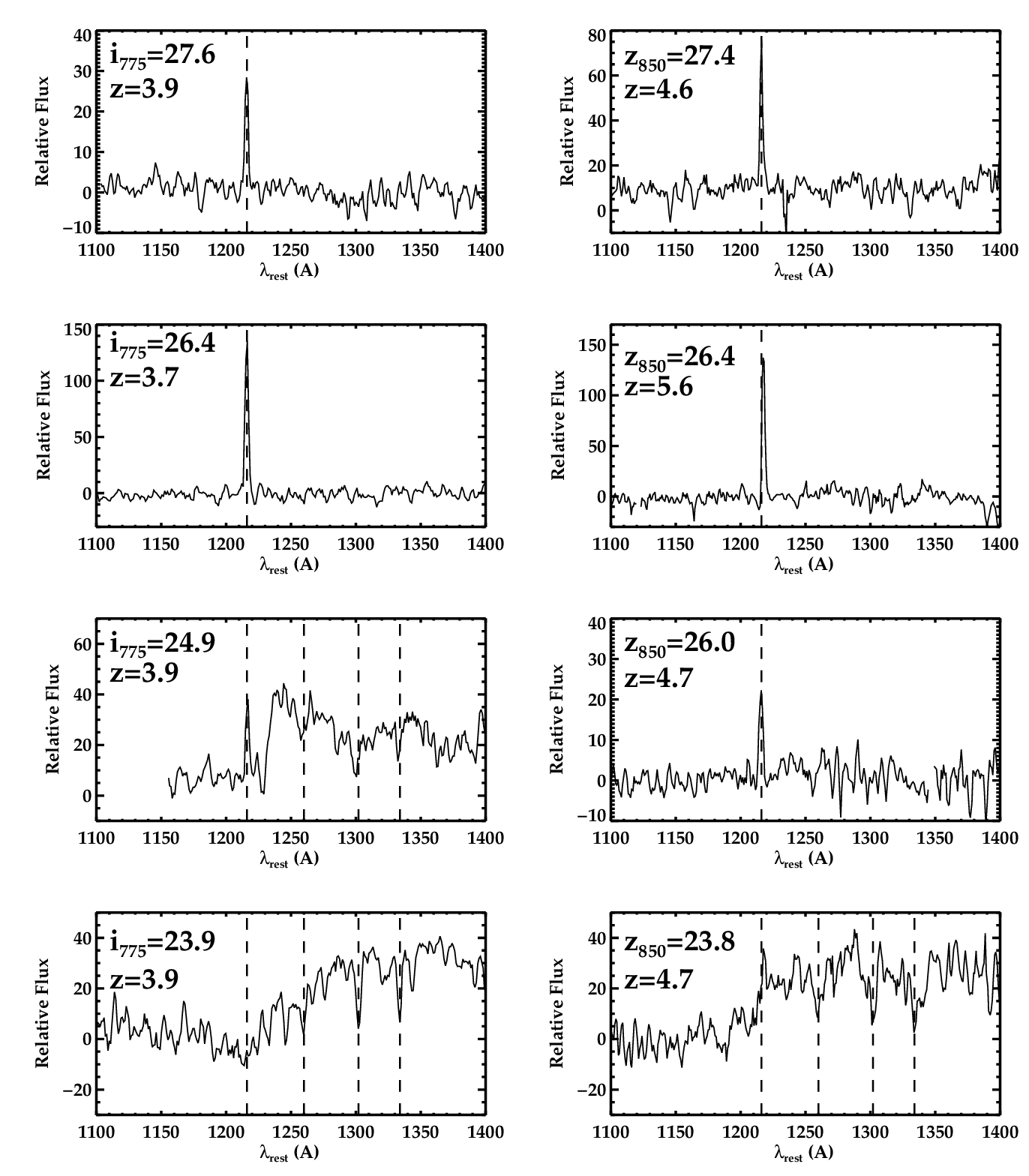}
\caption{Montage of DEIMOS 1D spectra from our survey arranged
  according to redshift and luminosity. B-drops are shown in the left
  column, and V-drops appear in the right column.  Each panel contains
  a label indicating the apparent magnitude of the continuum and
  redshift. Spectra are smoothed to a spectral pixel size of $\simeq
  3$~\AA. Where present, dashed vertical lines denote Ly$\alpha$, Si
  II 1260, OI+SiII 1303, and CII 1334 lines.}
\label{fig:deimos_spec}
\end{center}
\end{figure*}

We observed a final slitmask toward GOODS-S on 19-20 October 2009.
For this  run, we prioritised $i'$-drops and newly-discovered
$z$-drops \citep{Oesch10,McLure10,Bunker10,Wilkins10} and therefore
opted for the higher resolution 830 line mm$^{-1}$ grating blazed  at
8640~\AA~ with order blocking filter OG550.  With this set-up, the
spectra  generally provide coverage between 6800 and 10100 ~\AA~
(allowing Ly$\alpha$ to in principle be  detected between $z$=4.5 and
7.3), with each spectral pixel spanning 0.46~\AA.  Skylines are
measured to have a FWHM of $\simeq 2.4$~\AA, a significant improvement
upon the resolution  obtained with the 600 line grating.  However,
poor seeing (average of 1.3 arcseconds) and fog made redshift
confirmation particularly difficult.   

All data were reduced using the spec2d IDL pipeline developed for the
DEEP2 Survey.  The final reduction provides two-dimensional (2D)
spectra and variance arrays, along with a one-dimensional (1D)
extraction at the expected position of the dropout.  Wavelength
calibration was typically obtained from Ne+Xe+Cd+HG+Zn reference arc
lamps.  In general, the final wavelength solution is accurate to
within $\simeq 0.1$~\AA.  Examples of reduced 1D-spectra are presented
in Figure \ref{fig:deimos_spec}.

We flux calibrated the spectra using spectroscopic standard stars
observed  in the 24-26 March 2009 observing run.  We tested the flux
calibration  derived from these standards using spectra of alignment
stars included on  the slitmask (observed in 2 arcsec $\times$ 2
arcsec boxes).  We compute  optical broadband magnitudes for the
alignment stars from the flux-calibrated spectra using the appropriate
filter transmission functions.  The magnitudes measured from the
spectra match those from the  ACS images to within a factor of $\simeq
2$.  We bootstrapped a flux calibration on spectra from observing runs
for which spectroscopic flux standards were not taken (03-05 April
2008) using the measured flux in alignment stars that are in common
between the 2008 and 2009 observing  runs.  

Using the flux calibrations, we compute our typical 1$\sigma$ flux
sensitivity across the DEIMOS spectra for the 600 line grating.  These
measurements predict that we should detect continuum at the 1$\sigma$
level for sources with $V\simeq 25.5$ (per spectral resolution
element  in 1D spectra which have been extracted over $\simeq
1$\arcsec) across much of the wavelength regime covered by the 600
line spectra.  This prediction is consistent with expectations based
on the optical magnitudes of sources that show continuum traces in the
spectra.  The average 10$\sigma$ limiting line flux is
1-1.5$\times$10$^{-17}$ erg cm$^{-2}$ s$^{-1}$ between 6500\AA~ and
9300\AA, although sky lines become much more common toward the red
side of the spectra.   In a later section, we present more detailed
simulations that reveal the completeness for emission lines of a given
flux and redshift.

For the most recent run (October 2009), the limiting sensitivity was
significantly worse than in previous runs due to the seeing and fog.
Based on the signal obtained from the bright alignment stars on the
mask (each of  which has known broadband magnitudes) and the signal
from our spectroscopic standard stars, we estimate that our continuum
1$\sigma$ sensitivity (computed per resolution element  with a spatial
extraction width  of $\simeq 1.5$\arcsec) was $\simeq
3.5-4.0\times10^{-19}$ erg cm$^{-2}$ s$^{-1}$~ \AA$^{-1}$ in between
sky lines (corresponding to a continuum magnitude of $i'\simeq
24.25$).   Assuming typical line widths, this limit translates into a
10$\sigma$ line flux limit of 2.5-4.0$\times$10$^{-17}$  erg cm$^{-2}$
s$^{-1}$ for Ly$\alpha$.  As most of the sources targeted in this run
were faint ($z_{850}>26$ for the $i'$-drops), this means that we are
only sensitive to very strong emission lines (the 10$\sigma$
rest-frame equivalent width limit is $\simeq 75$-100 ~\AA~ for sources
with $z_{850}\simeq 26.5$).  For fainter  sources (e.g. the majority
of WFC3 $z$-drops in the UDF), the equivalent width limits are too
large to enable detection of Ly$\alpha$.

To summarise the current status of our DEIMOS observations, we have
obtained 549 spectra of B, V, $i'$, and $z$-drops.  To boost the S/N
of our spectra, many dropouts were observed on multiple masks; hence,
each spectra does not correspond to a unique sources.  Accounting for
this, we observed a total of 268 B-drops, 95 V-drops, 19 $i'$-drops,
and 17 $z$-drops.  We combine this sample with archival Keck and VLT
spectra in the  following two subsections, and in \S2.4, we present
the absolute magnitude  distribution of the combined VLT and Keck
spectroscopic samples.

\subsection{Archival Keck Spectroscopy in GOODS-N/S}

Additional $i'$-drops were observed with Keck/DEIMOS between 2003 and
2005 (Table 1).  Early observations were presented by Bunker et
al. (2003) and Stanway et al (2004), and an updated discussion,
including additional DEIMOS observations from 2003-2005, is given by
Bunker et al. (2010). These sources were observed  with the  1200 line
mm$^{-1}$ (described above) with seeing of 0\farcs 7 - 1\farcs 0.

In total, 45 $i'$-drops were observed in the two GOODS fields on 4
separate  slitmasks.  Of the 45 sources observed, 12 were included in
the  \citet{Vanzella09} VLT/FORS observations discussed below.
Integration times ranged between 2.3 and 10.5 hrs.  The $z_{850}$-band
magnitudes of the sources ranged between 24.7 and 28.3.  Given the
considerable range of exposure times, we take care  to estimate the
equivalent width completeness for sources of different  magnitudes on
the various masks that were observed. 

\subsection{Archival VLT/FORS Spectroscopy in GOODS-S}

Two programmes aimed at following up high-redshift  dropouts in
GOODS-S have been conducted with the VLT.  The first of these  used
the FORS2 multi-object spectrograph (Vanzella et al. 2005, 2006, 2008,
2009), and the second used the VIMOS multi-object spectrograph
\citet{Balestra10}.  Both teams have released their datasets to the
public.  In this paper, we focus on the FORS2 survey, as its survey
characteristics (resolution, spectral coverage) are closest to the
DEIMOS  survey.  We discuss the basic FORS2 survey details  below.

Between September 2002 and October 2006, the VLT FORS2 multi-object
spectrograph was used to observe sources identified in the GOODS
imaging  of CDF-S.  In total, 38 FORS2 masks were obtained using the
300I grism without an order-separating filter. Each mask was  observed
for roughly 4-6 hours.  In general, the spectra provide coverage
between 6000~\AA~ and 10000~\AA, with a  spectral resolving power of
R=660 which  provides resolution of $\simeq 13$~\AA~ at 8600~\AA.  

The FORS2 database provides redshift classifications and quality
grades (ranging from A to C) for the entire spectroscopic sample.  The
colour criteria used to select the dropouts in the FORS2 sample are
discussed in \citet{Giavalisco04b}; however, for the B-drops, a slight
variation  in the colour-selection was adopted  \citep[for details
  see][]{Vanzella09}.

In general, these criteria are very similar to those we have  adopted
for our DEIMOS survey.  Using the coordinates provided in  the public
FORS2 database, we query the version 2.0 ACS catalogs for GOODS-S  and
measure optical magnitudes in an identical fashion as described in
\S2.1.  Adopting the selection criteria used for the DEIMOS sample
(see  Stark et al. 2009), we find 83 B-drops, 56 V-drops, and 56
$i'$-drops that  were observed with FORS2.  Redshifts  were obtained
for 48 B-drops (46 with $z>3$), 37 V-drops (32 with $z>4$), and 26
$i'$-drops (21 with $z>5$).  

\begin{figure}
\includegraphics[width=0.47\textwidth]{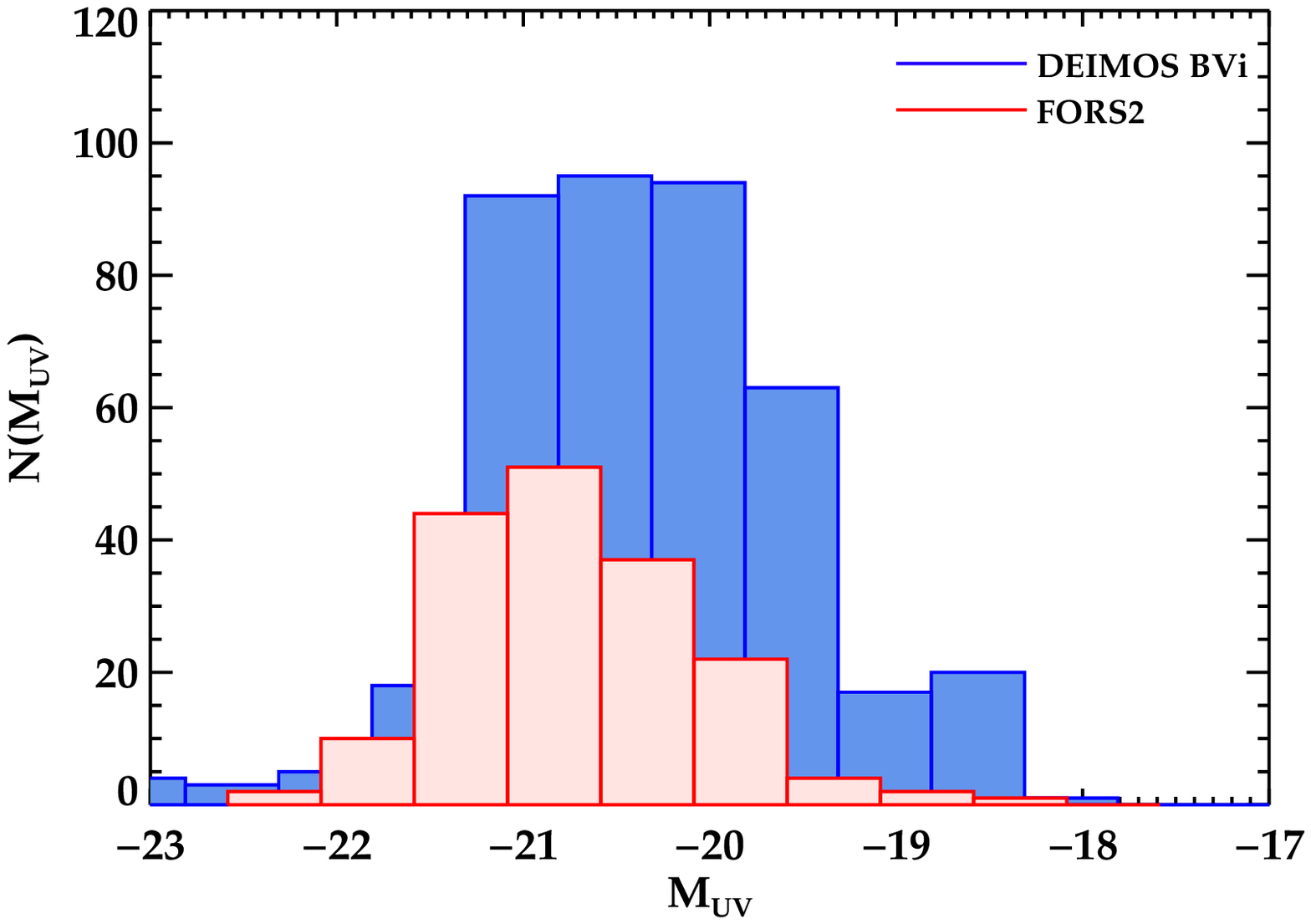}
\caption{Absolute magnitude distribution of FORS and DEIMOS
  spectroscopic samples.  The new DEIMOS B, V, and $i'$-drops are
  shown in shaded blue, and below the FORS2 observations
  \citep{Vanzella09} are shown in light red.  The DEIMOS datasets
  comprise the majority of the  spectra considered in our analysis.}
\label{fig:fors_v_deimos}
\end{figure}

The magnitude distribution of the FORS2 sample is generally weighted
toward  brighter sources, with few B and V-drops with magnitudes
fainter than  $z_{850} \simeq 26$.  Given the inherent faintness of
$i'$-drop samples, the magnitude distribution is significantly fainter
than for the lower redshift dropout samples, with $z_{850}\simeq
25-27$.  In Figure \ref{fig:fors_v_deimos}, we plot a comparison of
the  absolute magnitude distribution  of the FORS2 and DEIMOS dropout
samples.   The current DEIMOS sample  contributes 76\% of the B-drops
and 63\% of the V-drops and crucially extends the spectroscopic
coverage to lower luminosities.  

\subsection{Final Spectroscopic Sample}

\begin{figure*}
\begin{center}
\includegraphics[width=0.47\textwidth]{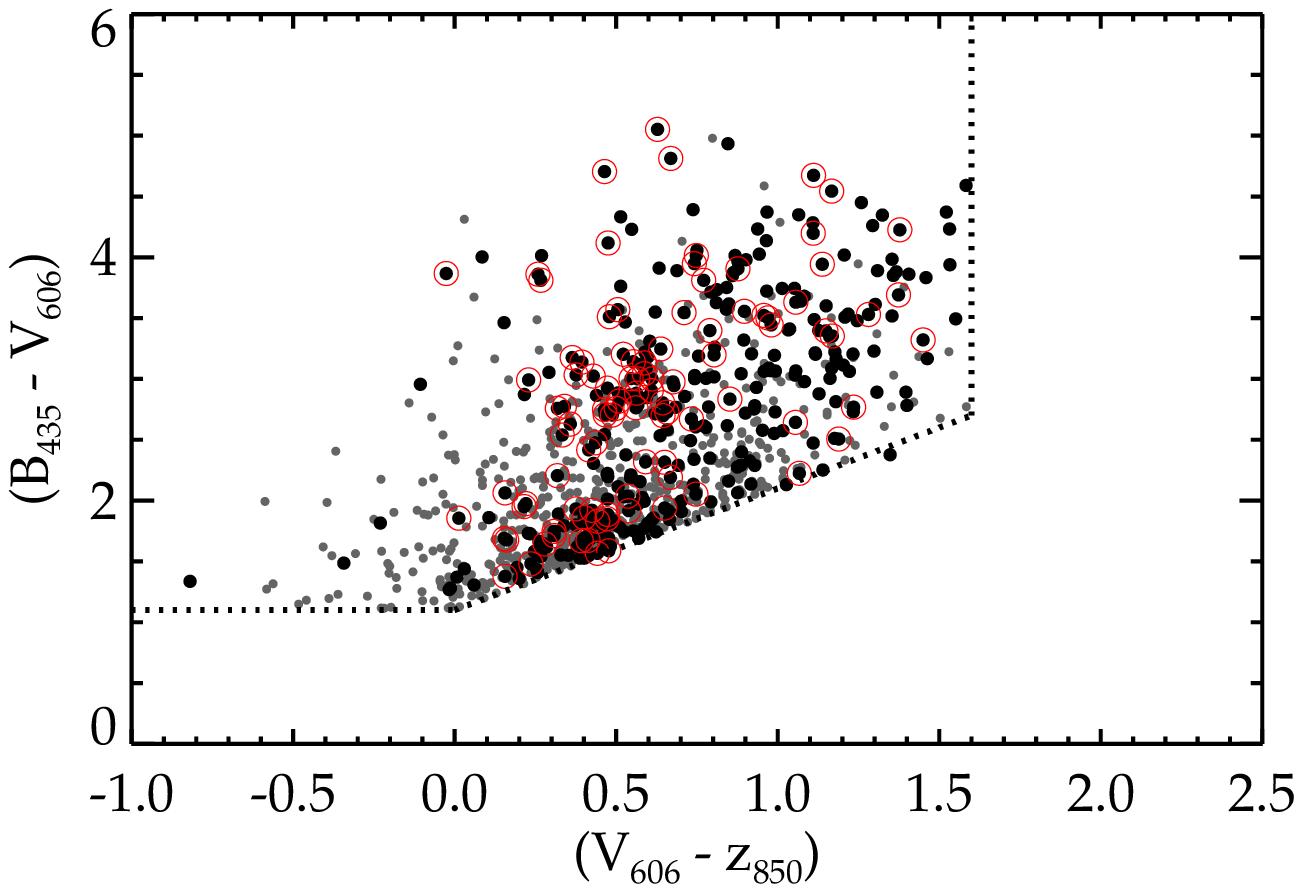}
\includegraphics[width=0.47\textwidth]{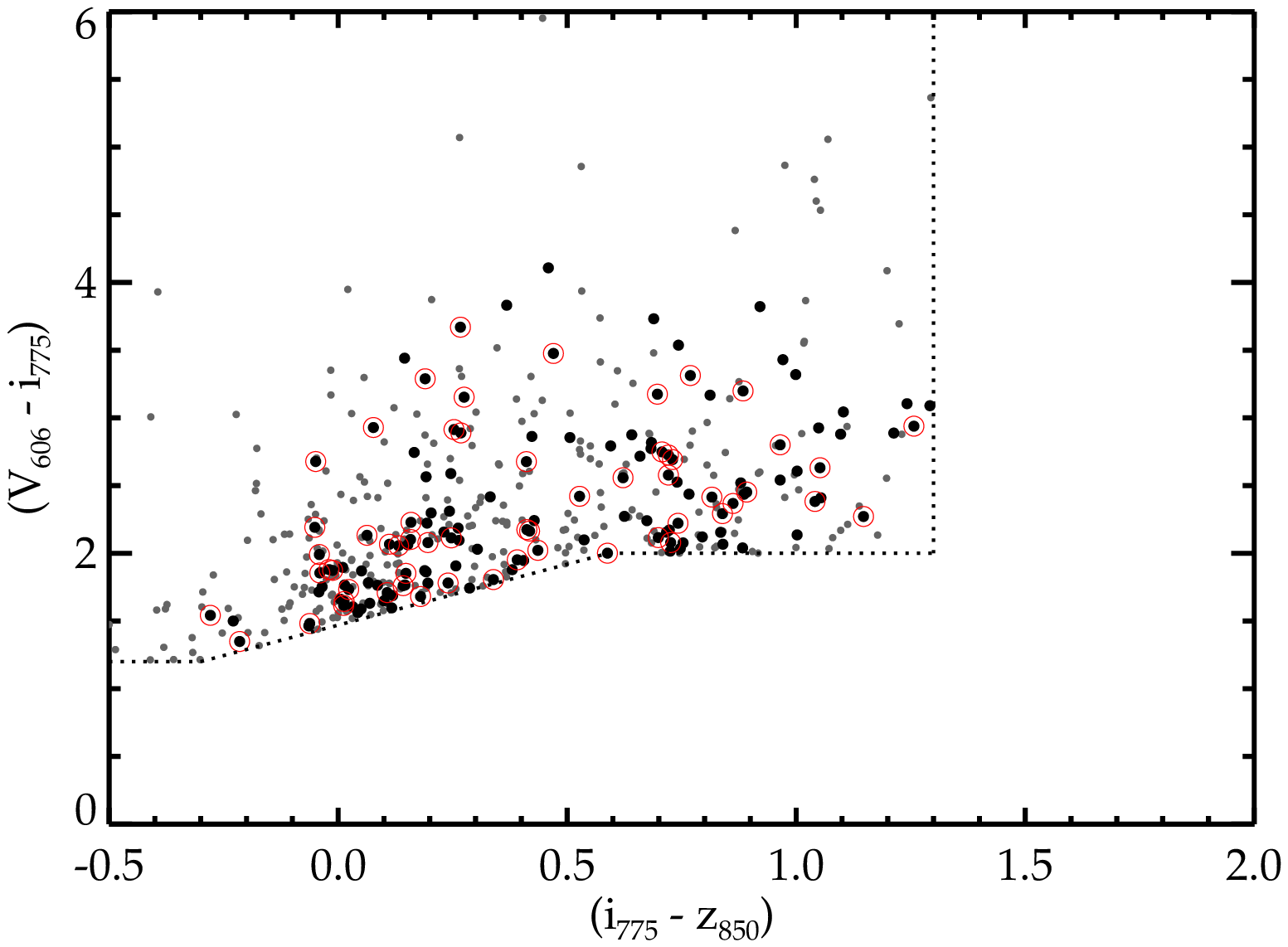}
\caption{Colour-colour diagrams for the B-drops (left) and V-drops in
  combined VLT and Keck spectroscopic surveys.  Objects with
  Ly$\alpha$ in emission are denoted with a second red circle.  The
  dotted lines show the  Lyman break selection criteria adopted in
  this paper.  The small grey circles show the distribution of colours
  for the  parent population of LBGs from \citet{Stark09}.}
\label{fig:bvdrops_color}
\end{center}
\end{figure*}

\begin{figure}
\includegraphics[width=0.47\textwidth]{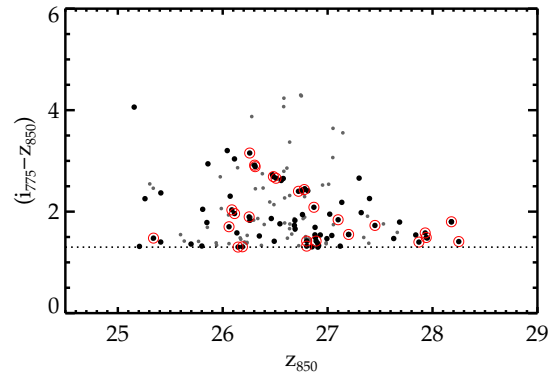}
\caption{$i'-z$ colour versus magnitude for $i'$-drops with
  spectroscopic observations with the dashed line showing the
  selection cut.   Points surrounded  by a red open circle indicates
  sources with Ly$\alpha$ seen in emission, and the small grey circles
  indicate the colours and magnitudes of $i'$-drops from the parent
  LBG samples \citep{Stark09}.}
\label{fig:idrops_color}
\end{figure}

When combined, the FORS2 and DEIMOS surveys contain a total of 627
spectra of unique high-redshift dropouts (351 B-drops, 151 V-drops,
108 $i'$-drops, and 17 $z$-drops).   In Figure
\ref{fig:bvdrops_color}, we present colour-colour diagrams of the
entire B and V-drop spectroscopic sample, indicating the selection
criteria.  The $i'$-drops were selected using the standard single
$i'-z>\rm{1.3}$ colour criterion (see Figure \ref{fig:idrops_color}).
Comparing the distribution of selected targets on these figures with
the larger sample of sources in the \citet{Stark09} photometric
catalog, we demonstrate that the colours of our targets with
spectroscopic coverage span nearly the full range of the parent LBG
population.  

\begin{figure}
\includegraphics[width=0.47\textwidth]{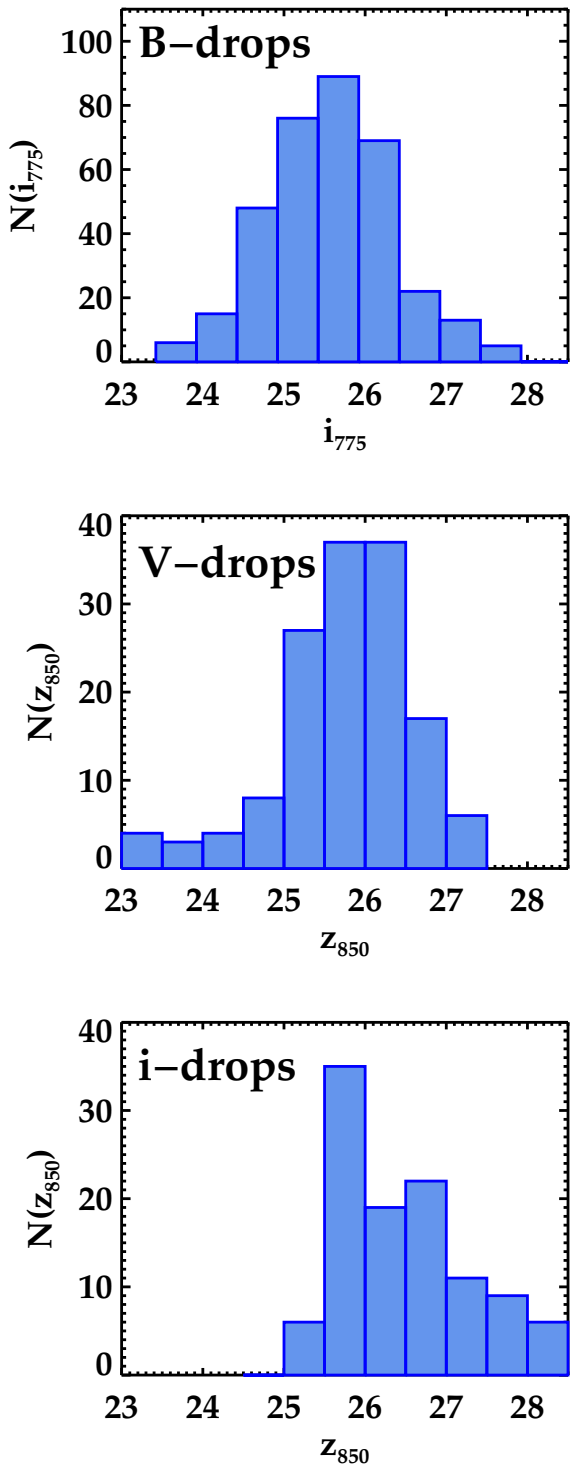}
\caption{Distribution of apparent magnitudes of B-drops (top), V-drops
  (middle), and $i'$-drops (bottom) in combined Keck and VLT
  spectroscopic  survey.  }
\label{fig:apparent_mag_dist}
\end{figure}

\begin{figure}
\includegraphics[width=0.47\textwidth]{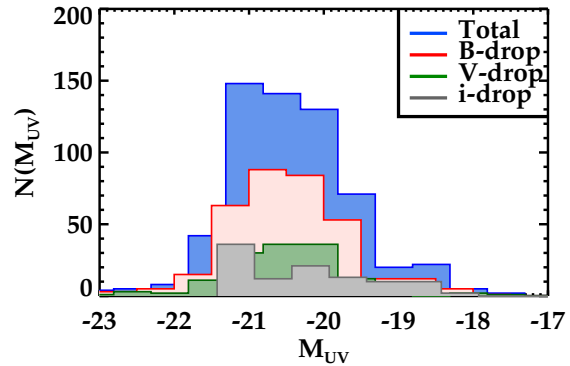}
\caption{Distribution of absolute magnitudes of sources in the
  combined  DEIMOS and FORS2 spectroscopic survey (shaded blue).  The
  shaded light red, green, and  grey denote the magnitude distribution
  of the  B, V, and $i'$-drop samples.  }
\label{fig:absolute_mag_dist}
\end{figure}

Figure \ref{fig:apparent_mag_dist} also shows the apparent magnitude
distributions of the various dropout samples.  While previous surveys
have focused on sources with apparent magnitudes brighter than $m
\simeq 26$, a major achievement of our survey is that we have been
able to push significantly below this limit with DEIMOS.   Our primary
motivation for doing so is that the characteristic UV luminosity
shifts to lower values at higher redshifts (e.g., Bouwens et al 2007), so to
compare  source properties over $3\lsim z\lsim 7$ in a meaningful
manner, probing deep becomes a necessity. To demonstrate our survey
has achieved this over $3<z<6$, in Figure \ref{fig:absolute_mag_dist},
we compare the absolute magnitude distribution of our various dropout
samples. Current estimates of the characteristic  absolute magnitude
at $z\simeq 7$ are  M$_{\rm{UV}}\simeq -19.9$
\citep{Bouwens09a,Ouchi09, Oesch10}.   As objects more luminous than
this are exceedingly rare at $z\gsim 6-7$, if we are to make
consistent Ly$\alpha$ fraction comparisons, we must probe to at least
this depth over $4<z<6$.

\begin{figure}
\includegraphics[width=0.47\textwidth]{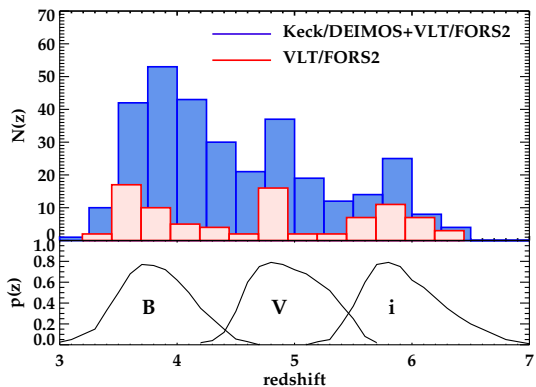}
\caption{Redshift distribution of B, V, and $i'$-drops with
  spectroscopic  confirmation in GOODS-N and GOODS-S (dark blue).
  This sample includes absorption  line systems in addition to the
  Ly$\alpha$ emitters discussed in this  paper.  The total sample
  contains 179 galaxies with redshifts in the  range $3\lsim z\lsim
  4.5$, 87 galaxies with redshifts over  $4.5\lsim z\lsim 5.5$, and 44
  galaxies with redshifts over $5.5\lsim z\lsim 6.5$.  For reference,
  we display  the redshift distribution of the VLT/FORS2 survey
  \citep{Vanzella09} in  light red.  }
\label{fig:z456dist}
\end{figure}

Redshifts were determined via visual examination of the spectra (see
Figure~ \ref{fig:z456dist} for a redshift distribution).  For UV faint
sources, the continuum is too faint for identification of the Lyman
break or absorption lines, so we can only measure spectroscopic
redshifts for those sources with Ly$\alpha$ in emission.  For UV
bright sources, redshift identification is performed via a combination
of the Lyman break, interstellar absorption lines, and/or Ly$\alpha$
emission.  We classify all redshifts according to their quality,
ranging  from A (definite), B (secure), C (possible/likely).  Nearly
all Ly$\alpha$  emitters fall into the first two categories owing to
the combination of  line profile, lack of other emission lines, and
strong continuum break (in spectra and/or imaging),  but some of the
absorption line detections are much more tenuous owing  to sky line
residuals and noise.  As the results of this paper are independent  of
the absorption line sample, we delay further discussion of the
absorption line catalog until a subsequent paper (Stark et al. 2010,
in  preparation), while details of the Ly$\alpha$ selection are
described  in the following section. However, we note that, in total,
the FORS2+DEIMOS sample contains 179 redshifts  over $3\lsim z\lsim
4.5$, 87 redshifts over $4.5\lsim z\lsim 5.5$, and  44 redshifts over
$5.5\lsim z\lsim 6.5$.

\section{Emission Line Measurements}

\subsection{Constructing the Ly$\alpha$ catalog}

Emission lines were initially identified visually in the
two-dimensional DEIMOS spectra and later in the one-dimension
extractions at the position of the dropouts.   We took care to
distinguish Ly$\alpha$ from other emission features (e.g. [OII] which
is resolved by DEIMOS) which correspond to lower-redshift galaxies.
Ly$\alpha$ emission is detected in 152 of the DEIMOS  dropouts.

For the VLT/FORS dataset, the redshifts were determined in
\citet{Vanzella09}.  As discussed in \S2.3, to ensure a uniform
selection across the Keck and VLT samples, we performed our own
photometric selection.   With the resulting  subset of 193 galaxies,
we identified those objects  showing Ly$\alpha$ in emission in their
one-dimensional spectra. 

We defined Ly$\alpha$ redshifts ($z_{Ly\alpha}$) for each object as
the  wavelength at which the Ly$\alpha$ line is at its peak flux
value.  As Ly$\alpha$ is generally redshifted by at least $\simeq 300$ km
s$^{-1}$ from  the frame of rest of the galaxy
\citep[e.g.][]{Shapley03,Vanzella09}, we note that this redshift is
not necessarily equivalent to the systemic redshift  of the galaxy.
Across both surveys, we identified Ly$\alpha$ emission in 108 B-drops,
63 V-drops, and 28 $i'$-drops.  The relative colour distribution of
these sources in relation to the larger spectroscopically-targetted
sample is shown in Figures \ref{fig:bvdrops_color} and
\ref{fig:idrops_color}.

\begin{figure}
\includegraphics[width=0.45\textwidth]{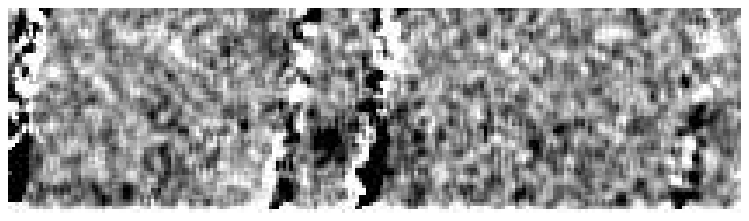}
\includegraphics[width=0.50\textwidth]{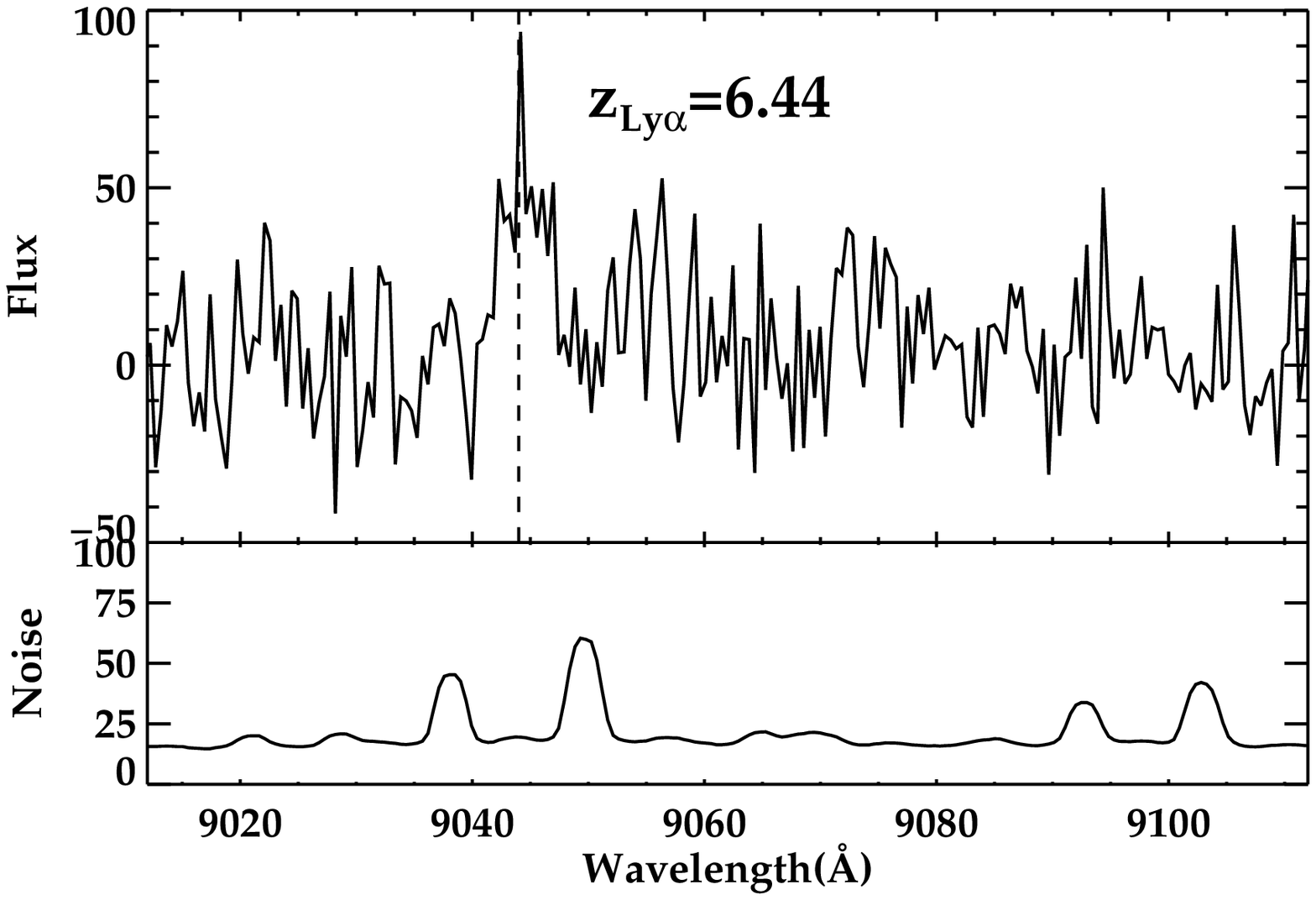}
\caption{Spectra of a J$\rm{_{125} \simeq 27.3}$ $z$-drop showing
  tentative Ly$\alpha$ emission at  9044~\AA.  {\it Top panel:} The 2D
  spectrum shows a significant emission feature detected  at the
  7$\sigma$ level  (black corresponds to positive flux in the image)
  in between two sky lines and centered exactly  at the position of
  $z$-drop '3' in \citet{Wilkins10}. {\it Bottom panel:} The 1d
  spectrum of the galaxy and associated  noise.  The spectrum and
  noise are not smoothed  to avoid blending with skylines at 9038~\AA~
  and 9048~\AA. The emission line between these features lies in a
  region of low noise, spanning $\simeq  10$~\AA~ in width.}
\label{fig:zdrop_spec}
\end{figure}

\begin{figure}
\includegraphics[width=0.50\textwidth]{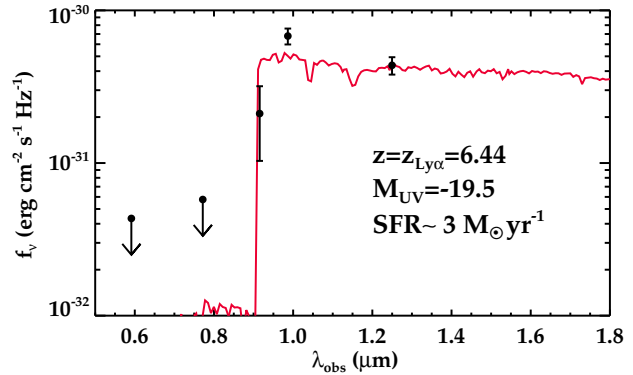}
\caption{Spectral energy distribution of $z$-drop '3' from
  \citet{Wilkins10}.  The datapoints show detections and upper limits
  from ACS and WFC3 imaging  of CDF-S.  Overplotted is a
  Bruzual-Charlot population synthesis model with  a redshift fixed at
  $z=6.44$, the spectroscopic redshift inferred from the  emission
  line detected in Figure \ref{fig:zdrop_spec}.  The broadband imaging
  data show a prominent break where the Lyman break is predicted and a
  significant excess of flux in the medium-band $\rm{Y_{098}}$ filter
  (relative to the $\rm{J_{125}}$-band), as expected in the presence
  of strong Ly$\alpha$  emission.}
\label{fig:zdrop_sed}
\end{figure}

In addition to the B, V, and $i'$-drops, we present tentative
spectroscopic  confirmation for one of the $z$-drops that was studied
in poor conditions in the October 2009 DEIMOS run (Table 1).  The only
$z$-drop to show a  clear emission feature at the expected object
location was source 3 in  \citet{Wilkins10}.  This object, identified
in the GOODS ERS WFC3 imaging of CDF-S (PI: O'Connell), is reasonably
faint (J$\rm{_{125} \simeq 27.3}$)  but is brighter than most of the
$z\gsim 6$ sources detected in the UDF
\citep[e.g.][]{Oesch10,Bunker10,McLure10}.   The DEIMOS spectrum
illustrates a significant 7$\sigma$ emission line in between sky lines
and centered at 9044~\AA~ (Figure \ref{fig:zdrop_spec}).  The S/N and
nearby sky line do not enable a robust measure of the asymmetry of the
line.   The  broadband SED is well fit by a source $z\simeq 6.44$ with
strong Ly$\alpha$ contaminating the medium-band $Y_{098}$ filter
(Figure~\ref{fig:zdrop_sed}).  If this emission feature is Ly$\alpha$,
it would lie at $z=6.44$, consistent with the photometric redshift
predictions.  Our flux calibration suggests  a flux of
$\rm{1.5\times10^{-17}}$ erg cm$^{-2}$ s$^{-1}$ and  hence an
equivalent width (rest-frame)  of 83~\AA.  We will confirm the redshift 
of this source in future campaigns of CDF-S.

\subsection{Computation of Ly$\alpha$ properties}

Next we compute the flux ($F_{Ly\alpha}$) and equivalent width
(W$_{Ly\alpha}$) for each Ly$\alpha$ emitter in our spectroscopic
sample.  Previous attempts to constrain evolution in the prevalence of
Ly$\alpha$ emitters have focused on measuring evolution in the
luminosity function of Ly$\alpha$ emitters (selected via narrowband
filters).  For our Ly$\alpha$ fraction test  described in \S1, we are
interested in determining the percentage of LBGs of a given luminosity
with Ly$\alpha$ emission much stronger than  the continuum flux.  This
test thus relies on accurate measurements  of the Ly$\alpha$
equivalent widths (W$_{Ly\alpha}$).

In principle, W$_{Ly\alpha}$  provides a more robust constraint than
$F_{Ly\alpha}$ as it does not rely on an absolute flux calibration.
However since the majority of spectra in our sample have broadband
continuum magnitudes fainter than the 1$\sigma$ continuum flux  limit
of our survey ($m\simeq 25-26$, \S2.1), we can only place an upper
limit on the continuum flux measured in this subset of our spectra,
thus producing a lower limit to the observed W$_{Ly\alpha}$.  We can
obtain a better estimate of the equivalent width by adopting  the
continuum flux measured from broadband imaging, taking care to  avoid
filters that are contaminated by Ly$\alpha$.  We detail the specific
procedure used to measure W$_{Ly\alpha}$ below.

For each spectrum, we compute the line flux, $F_{Ly\alpha}$, by
summing emission in excess of the continuum  between 1213 and
1221~\AA, taking care to avoid contribution from nearby sky lines or
artifacts and ensuring that the spatial extraction box covers the
entire spatial width of  the line emission. 

The measurement of the continuum flux depends on whether or not
continuum is detected.  For those objects with continuum detections,
we compute the average continuum  level redward of Ly$\alpha$,
$c_{red}$, by averaging the flux in regions between  OH sky lines
between 1225~\AA~ and 1255~\AA~ in the rest-frame of the galaxy.  As
discussed in \citet{Kornei09}, this should minimize contribution from
nearby absorption features.  The observed-frame W$_{Ly\alpha}$ is then
computed by taking the ratio of $F_{Ly\alpha}$ and $c_{red}$.  

For those objects without continuum detections, we compute equivalent
widths using the continuum level just redward of Ly$\alpha$ with the
broadband photometry discussed in the previous section and taking care
not  to include the contribution of Ly$\alpha$  to the continuum.  For
the B-drops and V-drops, we use the $i'_{775}$ and $z_{850}$-band
fluxes, respectively.  These measurements provide the luminosity at
rest-frame $\simeq 1500$~\AA, rather than just redward of Ly$\alpha$.
The  vast majority of sources are very blue (and hence nearly flat in
f$_\nu$), requiring no correction between the flux at 1500~\AA~ and
that at 1240~\AA~ (the central value in the wavelength-region summed
for continuum measurements  above).  For those few sources with very
red UV slopes (measured from the broadband  SEDs presented in Stark et
al. 2009), we apply a small adjustment to  correct for the change in
flux between 1500~\AA~ and 1240~\AA.  

For galaxies with redshifts above $z\simeq 5.7$, the $z_{850}$-band
filter  is contaminated by both Ly$\alpha$ emission, if emission is
present, as well as Ly$\alpha$ forest absorption shortward of rest
1216~\AA.   Accurate  continuum determinations must account for this.
If sources contain Y or J-band detections via WFC3 or ISAAC, then we
use  these measures for the UV continuum   However, at the moment, the
majority of UV-faint $i'$-drops are not detected with significance  in
the near-IR, so we must use $z_{850}$ detections for these sources.
For galaxies with no detected Ly$\alpha$ emission, we must only
correct for the Ly$\alpha$ forest absorption, which we do assuming the
redshift is identical to its value inferred from BPZ and that its SED
is flat in f$_\nu$.   Corrections range from $\simeq 0.2$ mag at $z=6$
to 0.7 mag  at $z\simeq 6.4$.  But for $z>5.7$ objects with  strong
Ly$\alpha$ emission, we also subtract off the line contribution when
computing the continuum flux.  Typical line contributions are $\simeq
0.1$-0.3 mag. 

\begin{figure}
\includegraphics[width=0.47\textwidth]{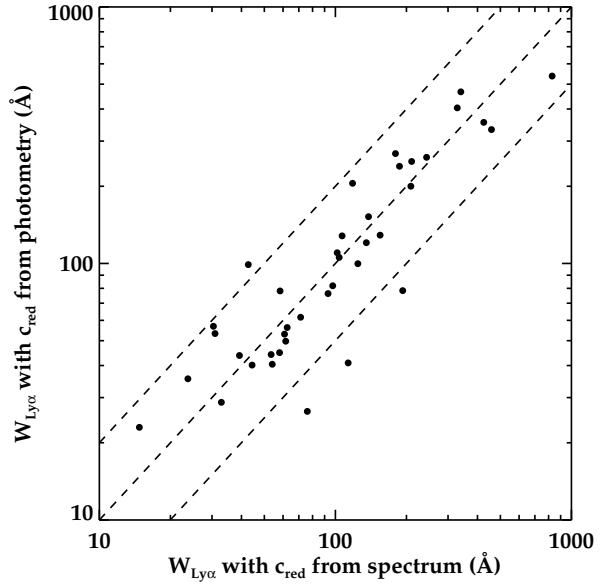}
\caption{Reliability of W$_{\rm{Ly\alpha}}$ measurements.  The x-axis
  shows equivalent widths (observed-frame) determined directly from
  spectra for sources  bright enough for a continuum measurement.  The
  y-axis shows the equivalent  widths measured for the same sources
  using the continuum flux from the broadband photometry.  The dashed
  lines demarcate values 0.5$\times$ and $2\times$ the
  W$_{\rm{Ly\alpha}}$ measured  directly from the spectrum. The median
  fractional  error from using equivalent widths derived using the
  continuum flux from broadband  imaging is $\simeq 20$\%.}
\label{fig:ew_comp}
\end{figure}

To test the reliability of the above method, we examine the subset of
galaxies  with bright UV-continua, comparing the continuum flux
determined  from broadband-imaging to that extracted from the spectra.
We find that the median fractional error using the estimate from
broadband imaging  is $\simeq 20$\% (Figure~\ref{fig:ew_comp}).  We
add this  error in quadrature to the photometric error on each
equivalent width  measurement.

\subsection{Completeness of Ly$\alpha$ detection}

In order to properly assess the fraction of Ly$\alpha$ emitters in our
sample as a function of UV luminosity and redshift, we must consider
how completeness varies with apparent magnitude and wavelength of the
Ly$\alpha$ emission line.  We estimate the typical completeness by
adding fake Ly$\alpha$ emission to random positions across the DEIMOS
and  FORS2 spectra.  In each case, we measure the line properties
($W_{Ly\alpha}$,  F$_{Ly\alpha}$, S/N) of the fake emission feature.
We ran enough trials (using a large number of spectra from each mask)
to obtain reliable estimates of the completeness of the Ly$\alpha$
recovery.  

\begin{figure}
\includegraphics[width=0.47\textwidth]{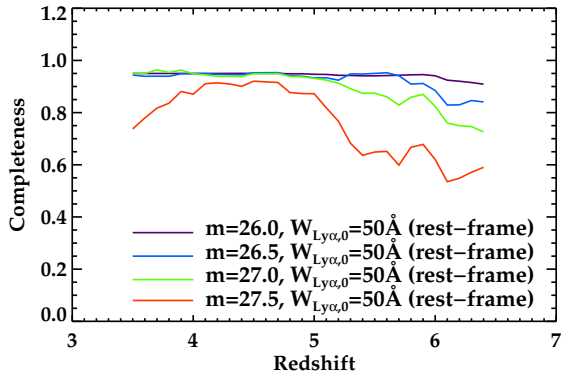}
\caption{Emission line recovery completeness.  The completeness  is
  determined via simulations placing fake lines in spectra and testing
  the rate at which sources about the equivalent width threshold  are
  recovered.  From top to bottom, the figure shows completeness versus
  redshift for $\rm{W_{Ly\alpha,0}}$ sources with continuum magnitudes
  (longward of the Lyman break) of 26.0, 26.5, 27.0, and 27.5.   }
\label{fig:completeness}
\end{figure}

Our goal is to identify a lower flux threshold above which Ly$\alpha$
is highly complete for most objects in our sample in order to minimize
the necessary completeness corrections.  We illustrate the results of
the completeness simulations for the 600 line mm$^{-1}$ grating on
DEIMOS in  Figure \ref{fig:completeness}.  The masks on this grating
included  B and V-drops which span the redshift range $3.5\lsim z\lsim
5.5$.  The simulations reveal that sources with $\rm{m_{AB}\simeq
  27.0}$  and with rest-frame Ly$\alpha$ equivalent widths
($W_{Ly\alpha,0}$)  in excess of 50~\AA~ are generally recovered over
nearly the entire redshift range probed by our dropouts (see Figure
\ref{fig:completeness}).  The high completeness for very strong
Ly$\alpha$ emitters amongst bright sources ($m\lsim 26$) arises as
these bright emission features correspond to fluxes that are in excess
of 10-20$\sigma$ and are recovered even when they lie on top  of sky
lines.  The $W_{Ly\alpha,0}>50$~\AA~ emission line completeness begins
to decline for very faint sources ($i'\simeq 27.5$) at the high-z tail
of the V-drop redshift distribution.  For the  FORS2 data, the spectra
show similarly high completeness for B and V-drops  with
$W_{Ly\alpha,0}>50$~\AA.  These simulations suggest that the measured
Ly$\alpha$ fraction  of sources in the redshift range $3.5\lsim z\lsim
5.5$ should not suffer from significant incompleteness for B and
V-drops more  luminous than $M_{\rm{UV}}\simeq -19$.  

For the $i'$-drops, the completeness in the deep DEIMOS and FORS2
spectra remains high for strong (e.g. $W_{Ly\alpha,0}>50$~\AA~)
Ly$\alpha$ lines.  However, it is clear from
Figure~\ref{fig:completeness}  that incompleteness is not negligible,
particularly for faint ($z_{850}>27.0$)  sources at $z\gsim 5.8$ which
are  less than 80\% complete.  We thus adopt this as our $i'$-drop
magnitude  threshold for inclusion into the Ly$\alpha$ fraction test,
limiting  us to sources more luminous than  $M_{\rm{UV}}\simeq -19.7$.
We also do not  include the 17 $i'$-drops on the mask GS051 (see Table
1) from Bunker et al. 2010 (in preparation) owing to its reduced
integration time of 2.3 hours.  

\subsection{Contamination in spectroscopic samples}

Although foreground emission line sources (e.g. [O II] emitters can be
readily distinguished from those revealing Ly$\alpha$ (\S3.1), our
Ly$\alpha$ fraction test requires that we have a reliable sample of
LBGs for which no emission line is seen.

While the Lyman break selection criteria are chosen to minimize the
inclusion  of low-redshift and stellar contaminants, it is clear  that
interlopers still populate dropout samples.
For our purposes, it  is important that we consider the luminosity and
redshift dependence of contaminants, as these could create artificial
trends in our derived Ly$\alpha$ fractions.   We expect low-$z$
contamination  to increase toward the fainter end of our sample.  This
is primarily due  to photometric scatter; since faint sources are
detected with lower S/N and  have less dynamic range available to
constrain the break, it is more likely  that a faint low-$z$ source
will be scattered into the LBG selection window.   However, if the
typical contamination is fairly low in all luminosity bins
(e.g. $\lsim 10$\%), then this would only require a minor correction
to the luminosity dependence of the Ly$\alpha$ fraction.

To investigate this in more detail, we compute photometric redshift
probability distributions for the spectroscopic sample using the
observed photometric catalogues compiled in \citet{Stark09}, updated
to include v2 GOODS ACS photometry, and the BPZ software (Benitez et
al. 2000).  Details of the photometric redshift methodology is
discussed  in \citet{Stark09}.  Using the probability distributions
derived from BPZ, we compute the probability that each object lies
outside the redshift range constrained by our spectroscopic
observations  (typically $z\lsim 3.4$).  We then place those galaxies
without spectroscopic  redshifts in bins of UV luminosity and compute
the total contamination fraction as a function of UV luminosity.   As
expected, the results show that the contamination fraction increases
toward lower luminosities.  We find negligible contamination for
bright sources ($\rm{-22\lsim M_{\rm{UV}}\lsim -20}$) with SEDs
constrained by high S/N photometry.  In the two faintest absolute
magnitude bins considered ($\rm{M_{\rm{UV}}=-19}$ and -18), the
predicted contamination increases to $\simeq 10$\%.   This suggests
that, owing to low-$z$ contamination, the true Ly$\alpha$ fractions in
the two faintest bins should be 1.11$\times$ greater than derived.  We
consider the effects of this in our discussion of the Ly$\alpha$
fraction in \S4.1.   

We now examine whether the SEDs predict contamination should vary
strongly with redshift.  As above, we measure the contamination
fraction (of objects that are not Ly$\alpha$ emitters) implied for  B,
V, and $i'$-drop samples from the photometric probability
distributions.  As above, we find negligible contamination at the
brighter magnitudes.  Combining sources with UV luminosities spanning
$\rm{-20\lsim M_{\rm{UV}}\lsim -18}$, we find that the contamination
fraction  increases from 2\% for V and $i'$-drops to 5\% for
B-drops. This suggests  that in this luminosity regime, the B-drop
Ly$\alpha$ fraction will be underestimated by a factor  of 1.05, while
the V and $i'$-drop Ly$\alpha$ fraction will be underestimated  by a
factor of 1.02.  Hence low-$z$ contamination will cause the positive
evolution in the Ly$\alpha$ fraction with redshift to be overestimated
by a factor of 1.03 in this luminosity range.  As we will show in
\S4.2,  while the redshift-dependent trends in the Ly$\alpha$ fraction
are small  in their amplitude, this contamination effect contributes
little to the observed  variation with redshift.

\section{Analysis}

We now turn to the key questions posed in \S1. Armed with a large
sample of LBGs for which some fraction, $x_{\rm{Ly\alpha}}$, show
Ly$\alpha$ emission, we discuss what can be learned about the
demographics of line emission in the  star-forming population and how
such trends might affect our ability to use Ly$\alpha$ as a tracer of
reionization. In this analysis section, first we present the results,
both in terms of the fraction of LBGs showing line emission as a
function of luminosity and redshift. We then discuss the physical
factors that might explain these trends prior to the use of Ly$\alpha$
emission statistics as a possible probe of reionisation.

\subsection{The Luminosity-Dependence of Ly$\alpha$ emission at high-redshift}

First, we discuss the relationship between  Ly$\alpha$ emission and
luminosity in our spectroscopic sample of Ly$\alpha$ emitters between
$3\lsim z\lsim 6$. If the Ly$\alpha$ fraction  varies strongly with
luminosity, as may be expected given recent claims of
luminosity-dependent  dust obscuration at high-redshift
\citep{Reddy09,Bouwens09b}, then care must be taken to compare only
galaxies of similar luminosity when searching for evolution in the
Ly$\alpha$ fraction with redshift. As our survey probes to
considerably lower luminosities than past spectroscopic LBG samples at
$z\gsim 3$ (M$_{UV}\simeq -18$, Figure~\ref{fig:absolute_mag_dist}),
our sample is well-suited to investigating such a relationship.

There have been a number of previous studies which examine how
Ly$\alpha$ line strength varies with luminosity.  Many of these
studies have reported a correlation between Ly$\alpha$ equivalent
width and UV luminosity.  \citet{Shapley03} binned their sample of
$z\simeq 3$ galaxies  with $W_{Ly\alpha,0}>20$~\AA~  in three groups
of apparent UV luminosity and found that the mean of the
$W_{Ly\alpha,0}$ distribution increased toward  fainter UV luminosity.
Others have examined Ly$\alpha$ equivalent widths as a function of UV
continuum luminosity, revealing a deficit in large equivalent width
Ly$\alpha$ lines in the most luminous continuum sources
\citep{Ando06,Ouchi08, Pentericci09,Vanzella09,Balestra10}.
Additionally,  by comparing the UV luminosity function of LBGs with
that of  narrowband-selected LAEs,  \citet{Ouchi08} has shown that
Ly$\alpha$ emitters are likely to be more  prevalent at the faint-end
of the luminosity function.  As we  discussed earlier, these results
are consistent with simple theoretical  expectations in which low
luminosity galaxies are less obscured by dust (due to lower
metallicities) and perhaps have  lower column densities (or covering
fraction) of HI surrounding them
\citep[e.g.][]{Verhamme06,Verhamme08,Schaerer08}. Perhaps of equal or
greater  importance is the bulk velocity field of the HI
\citep{Shapley03, Steidel10}, which  we discuss further in \S4.3.

However, others have found the evidence for  a correlation between
luminosity and Ly$\alpha$ equivalent width to be less convincing
\citep[e.g.][]{Steidel00,Nilsson09}.  In particular, \citet{Nilsson09}
demonstrated that the dearth of luminous galaxies with extreme
Ly$\alpha$  emission in magnitude or flux-limited surveys does not
require the  equivalent width distribution to be luminosity-dependent.
The  lack of such extreme emitters is actually expected in a magnitude
or flux limited  survey, as the most luminous and the most extreme
emitters are both  rare, causing this portion of parameter space to
poorly represented unless very large volumes are covered.  This is an
important realisation but only considers LBGs with continuum
luminosities  brighter than M$_{\rm{UV}}\simeq -20$, significantly
more luminous than the feeble  sources probed in our survey.
Likewise, \citet{Kornei09} found only marginal evidence for a
correlation between $W_{Ly\alpha}$ and UV continuum luminosity in a
large sample of $z\simeq 3$ LBGs, but concluded that this may result
from the limited dynamic range in UV luminosity probed by their
sample.  

\begin{figure}
\includegraphics[width=0.47\textwidth]{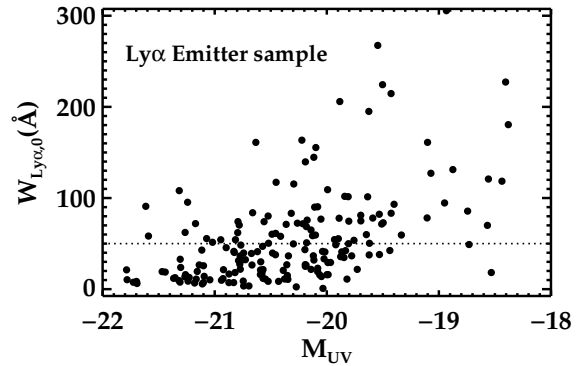}
\caption{Rest-frame Ly$\alpha$ equivalent width
  (W$_{\rm{Ly\alpha,0}}$)  as a function of rest-frame  absolute
  magnitude, M$_{\rm{UV}}$ for Ly$\alpha$ emitters in our
  spectroscopic  sample.  The sample is nearly complete at all
  M$_{UV}$ for sources with W$_{\rm{Ly\alpha,0}}>50$~\AA~ (dotted
  line).  }
\label{fig:wlya_dist}
\end{figure}

Figure~\ref{fig:wlya_dist} shows the distribution of rest-frame
equivalent width, W$_{Ly\alpha,0}$ for Ly$\alpha$ emitters in our
survey as a function of rest-frame UV luminosity, $M_{\rm{UV}}$.  In
considering this figure we must choose a sufficiently bright
W$_{Ly\alpha,0}$  threshold to avoid incompleteness in the faintest
sources.  As demonstrated in \S3.3 and Figure~\ref{fig:completeness},
sources with W$_{Ly\alpha,0}$ in excess of 50~\AA~ are detected with
high completeness  ($>90$\%) across our spectra, so we adopt this
value as our equivalent  width threshold.  The data show an apparent
lack of strong line emission among the most luminous dropouts, as has
been found elsewhere \citep{Ando06,Pentericci09,Vanzella09}, yet as
demonstrated in \citet{Nilsson09}, this does not necessarily imply
that Ly$\alpha$ emission is less common in luminous galaxies.
However, when  Figure~\ref{fig:wlya_dist} is viewed in concert with
the absolute magnitude histogram of our spectroscopic sample (Figure
\ref{fig:absolute_mag_dist}), it becomes clear that strong line
emission must be more common in low luminosity dropouts, for the
number of low luminosity galaxies ($M_{\rm{UV}}>-19.5$) targeted  in
our campaign is as low as the most luminous sources
($M_{\rm{UV}}<-21.5$), but the number of strong line emitters is far
greater among the feeble sources.  We emphasize that owing to our
equivalent width threshold of W$_{Ly\alpha,0}\simeq 50$~\AA, the
Ly$\alpha$ emitters amongst low  luminosity sources are secure and
robust ($S/N>10$) detections (similar  to those at the top of Figure
1); hence this result is not driven by noisy features. 

In order to most clearly quantify the luminosity-dependence of
Ly$\alpha$ emission, we must compute the {\it fraction of LBGs in our
  spectroscopic sample that show strong  Ly$\alpha$ emission},
$x_{\rm{Ly\alpha}}$,  as a function of emerging UV luminosity.  Given
that the majority of our spectroscopic sample is fainter than our
continuum  flux sensitivity (Figure~\ref{fig:apparent_mag_dist}), it
is  not possible to measure redshifts for  faint sources without
emission.  Hence if we were to compute x$_{\rm{Ly\alpha}}$ based soley
on sources with confirmed redshifts (as has been done previously  for
brighter samples), we would artificially increase $x_{\rm{Ly\alpha}}$
toward lower luminosities.  To avoid this bias, we define
$x_{\rm{Ly\alpha}}$ as the number of Ly$\alpha$ emitters above some
W$_{Ly\alpha,0}$ threshold divided by the total number of dropouts
placed on our slitmasks.  

The error on $x_{\rm{Ly\alpha}}$ is computed as follows.  We first
derive the Poisson error from the number of sources considered in each
luminosity or redshift bin.  In addition to the Poisson error, each
measurement is subject to additional error owing  to uncertainty in
the equivalent width measurements.  We thus conduct Monte Carlo
simulations, randomly varying the equivalent width of each galaxy
assuming a normal distribution with mean and standard deviation
corresponding  to the measured W$_{\rm{Ly\alpha,0}}$ and $\sigma_{W}$.
For each realization,  we compute the luminosity-dependent Ly$\alpha$
fraction.  Considering all the trials, we compute the standard
deviation in the x$_{Ly\alpha}$ distribution.  We then combine this
error term in quadrature  with the random error derived above.  In
doing  these simulations, we also consider whether equivalent width
error may artifically scatter a net flux of feeble galaxies above our
equivalent width threshold, introducing the observed trend.  However
the  simulations demonstrate that this is not the case, as the trend
is readily  apparent in each realisation.  

In constructing the Ly$\alpha$ fraction, we must also recognize that
the wavelength  coverage of some spectra is such that Ly$\alpha$ would
not be recovered across the full redshift range over which those
dropouts might be selected.  This applies mostly to FORS2 spectra of
B-drops which can only detect Ly$\alpha$ for sources with $z\gsim
3.9$.  Given that B-drops are expected to span the redshift range $3.5
\lsim z\lsim 4.5$, this suggests that a large number of B-drops would
not be recovered in the FORS2 spectra even if they showed Ly$\alpha$
in emission.  We therefore limit our study of B-drops to those with
spectra that allow Ly$\alpha$ to be detected over the entire redshift
range  expected.  This is a particular advantage of the DEIMOS
component of our survey where the numerous B-drops taken with  the 600
line mm$^{-1}$ grating are fully sampled. 

\begin{figure*}
\includegraphics[width=0.47\textwidth]{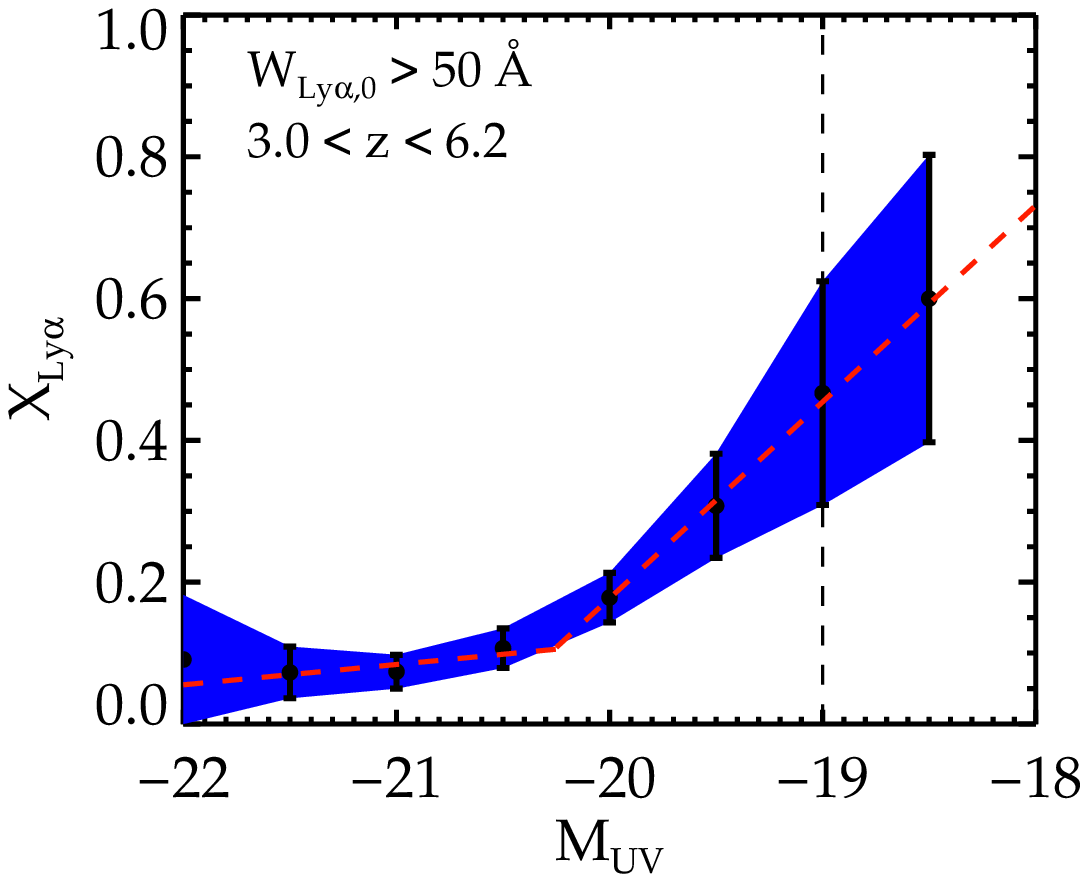}
\includegraphics[width=0.47\textwidth]{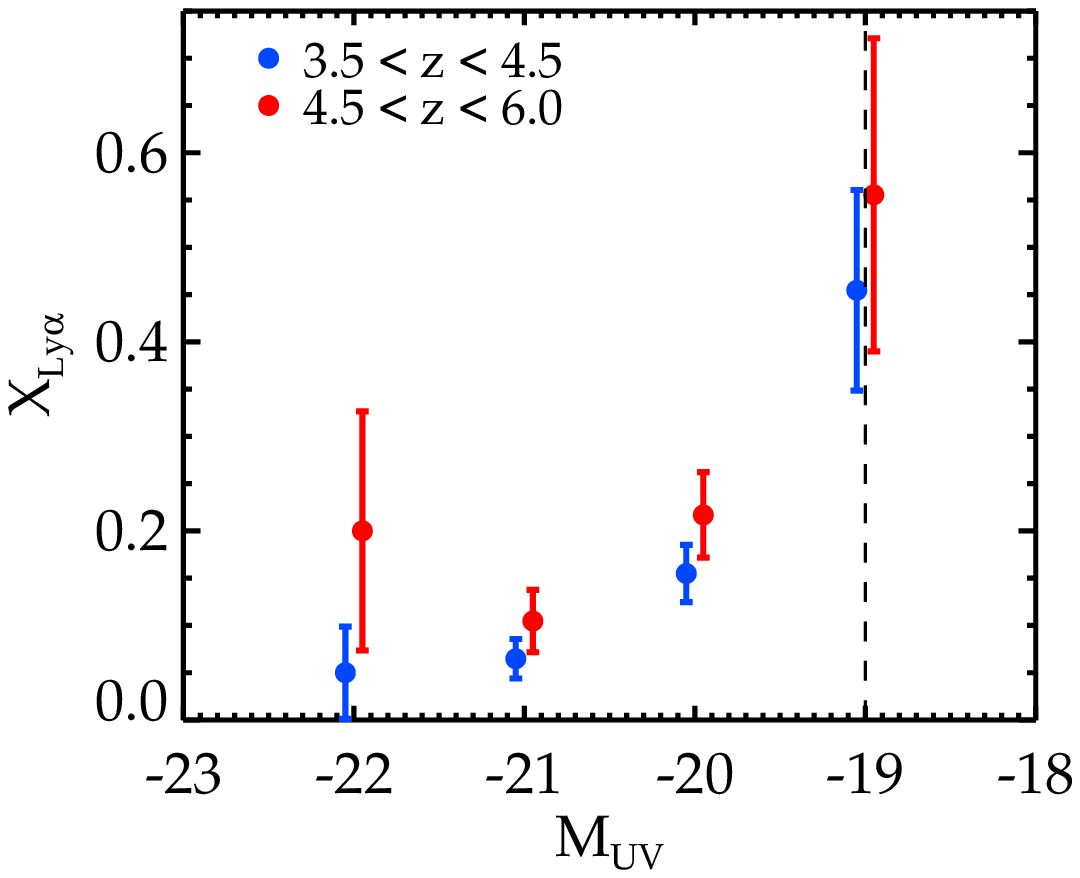}
\caption{({\it Left:}) Fraction of spectroscopic dropout sample
  showing strong Ly$\alpha$ emission (W$_{\rm{Ly\alpha,0}}> 50$~\AA)
  as a function of UV luminosity.  The dashed lines corresponds to
  first order polynomial fits to the Ly$\alpha$ fractions in the range
  $-22.0\lsim M_{\rm{UV}}\lsim -20.5$ and  $-20.0\lsim
  M_{\rm{UV}}\lsim -18.5$.  ({\it Right:}) Fraction of spectroscopic
  dropout sample showing strong Ly$\alpha$ emission
  (W$_{\rm{Ly\alpha}} > 50$~\AA) as a function of UV luminosity for
  samples with $3.5<z<4.5$ (blue circles) and $4.5<z<6.0$ (red
  circles).  Vertical lines correspond to $\simeq 90$\% completeness
  limits for B and V-drop samples.    For the $i'$-drops, the
  completeness limits are $\simeq 0.7$ mags brighter.  Any
  incompleteness would serve to increase the Ly$\alpha$ fractions
  further.  } 
\label{fig:lya_frac}
\end{figure*}

Following this procedure, we compute the luminosity-dependence of our
entire spectroscopic sample.  We find that $x_{\rm{Ly\alpha}}$ is
considerably larger in  low luminosity galaxies (Figure
\ref{fig:lya_frac}), increasing from 10-20\%  for luminous sources
(M$_{\rm{UV}}\simeq -21$) to 60-70\% for feeble galaxies
(M$_{\rm{UV}}\simeq -18-19$).  A similar trend is seen if we adopt
larger equivalent width thresholds.  The Ly$\alpha$ fraction rises
slowly with decreasing  luminosity over $-22.0\lsim M_{\rm{UV}}\lsim
-20.5$ but then begins to increase  more rapidly at lower
luminosities( $-20.5\lsim M_{\rm{UV}}\lsim -18.5$).  We thus fit first
order polynomials over each of these luminosity ranges,  finding
$x_{\rm{Ly\alpha}} = \rm{1.09 + 0.047~M_{\rm{UV}}}$ at the luminous
end and $x_{\rm{Ly\alpha}} = \rm{5.46 + 0.26~M_{\rm{UV}}}$ for the
lower luminosity sources.   Given that the Ly$\alpha$ fraction
increases slowly for luminous sources, it is perhaps no surprise that
previous studies which were limited to this luminosity regime found
only moderate evidence for luminosty  dependent trends
\citep{Nilsson09,Kornei09}; it is only by probing lower luminosity
galaxies that we start to see a clear trend in the prevalence of
Ly$\alpha$ with luminosity.

Intriguingly, the ``break'' in the luminosity-dependence of the
Ly$\alpha$  fraction in Figure \ref{fig:lya_frac} occurs at very
similar luminosities  to the characteristic UV luminosity of the
rest-frame UV $4<z<6$ luminosity  function (e.g.,
\citealt{Bouwens07}). At luminosities greater than L$^\star_{UV}$, the
Ly$\alpha$ fraction is low and increases slowly, but below this
luminosity, the  Ly$\alpha$ fraction increases to much larger values.
While perhaps coincidence, this may  suggest that whatever process
modulates the knee of the  luminosity function at $z>4$ may play a
role in the escape of Ly$\alpha$ photons.

In order to put these results in context, it is interesting to
estimate  the escape fraction of Ly$\alpha$ photons that is implied by
these  large equivalent widths.  In particular, we are interested  in
the Ly$\alpha$ escape fraction that is implied by the
$W_{Ly\alpha,0}\simeq 50$~\AA~ threshold we have adopted.  Assuming a
Salpeter IMF with M$_{\rm{upper}}$=120 M$_\odot$, case B
recombination,  constant star formation, and  metallicity ranging
between $Z$=1/20 Z$_\odot$ and Z$_\odot$, Ly$\alpha$  equivalent
widths can be as high as $\simeq 200-300$~\AA~ in the first few  Myr
of the star formation episode, asymptoting to $\simeq 100$~\AA~ after
10 Myr (\citealt{Malhotra02,Schaerer03}).  Hence our equivalent width
threshold corresponds to a Ly$\alpha$ escape fraction of 15-50\%.  Of
course, this assumes that all ionising  photons are absorbed; if there
is significant Lyman continuum leakage,  then the predicted equivalent
widths would decrease, increasing the  inferred Ly$\alpha$ escape
fraction.  Alternatively, if dust is confined to cold neutral clouds,
then the maximum Ly$\alpha$  equivalent widths may be larger than
quoted above \citep{Neufeld91,Hansen06,Finkelstein08},  decreasing the
implied Ly$\alpha$ escape fraction.

These results provide clear evidence that Ly$\alpha$ emission becomes
continuously more prevalent among lower luminosity star forming
galaxies.  Indeed, it appears that the {\it majority} of feeble
(M$_{\rm{UV}}<-19$) sources are  strong Ly$\alpha$ emitters.  These
results appear to indicate that the escape fraction of Ly$\alpha$
photons (relative to that of far-UV continuum photons) is strongly
luminosity-dependent and at very low continuum luminosities may
commonly exceed the 5\% robustly derived (at $z\simeq 2$) via
Ly$\alpha$  and H$\alpha$ surveys \citep{Hayes10}.  In the following
section, we examine whether similar trends are seen with redshift,
while in \S4.3 we examine the physical mechanism is that are likely to
be governing the escape fraction  of Ly$\alpha$ photons at $z\gsim 3$.

\subsection{Variation in the Ly$\alpha$ fraction in the redshift 
range $3\lsim z\lsim 6$}

We now examine the redshift evolution of the prevalence of Ly$\alpha$
emitters in the Lyman break galaxy population over $3<z<6$.  Since the
IGM appears to be highly ionised over this redshift range, this
measurement  provides the opportunity to understand the extent to
which  factors other than IGM  attenuation affect the Ly$\alpha$
fraction.  By  calibrating these effects over this redshift interval,
we can more  accurately detect the signal of reionisation on the
Ly$\alpha$ fraction.

We consider luminosity-dependent samples in two separate redshift
bins, $3.0<z<4.5$ and $4.5<z<6.0$ (Figure~\ref{fig:lya_frac}). Sources
without spectroscopic redshifts are placed into one bin or the other
based  on their photometric redshift (see \citealt{Stark09} for
discussion of  photometric redshifts).  In each UV luminosity bin, the
fraction $x_{\rm{Ly\alpha}}$ increases with redshift.  The two bins
with the lowest error and incompleteness ($M_{\rm{UV}}\simeq -21$ and
-20) show increases of 60\% and 40\%, respectively.   Our equivalent
width threshold and redshift binning ensures  that this result is not
biased by redshift-dependent incompleteness.  To determine the
differential growth, we compute the average change in Ly$\alpha$
fraction,  $\Delta x_{\rm{Ly\alpha}}$, across all luminosity  bins
(weighted by the inverse variance of each bin) and compute $\Delta z$
using the median redshift in each of the two bins.  With this
approach, we find that the Ly$\alpha$ fraction increases with redshift
following $\rm{dx_{Ly\alpha}/dz \simeq 0.05}$.  As we mentioned in
\S3.4, low-$z$ contamination appears to decrease very slightly  with
redshift.  However, we find that this effect introduces such small
changes in the Ly$\alpha$ fractions ($\lsim 1$\%) in each redshift
bin,  such that the weighted Ly$\alpha$ fraction redshift evolution
remains as  quoted above.  Hence Ly$\alpha$ fractions grow by nearly
$\Delta x_{\rm{Ly\alpha}} \simeq 0.1$ at fixed $\rm{M_{\rm{UV}}}$
between $z\simeq 4$ and $z\simeq 6$.   

We emphasize that this trend is not driven by biases associated with
LBG selection.  It is well-established that the presence of strong
Ly$\alpha$ can affect the broadband colours used for dropout selection
\citep{Stanway07,Stanway08}.  Line emission can either boost the
dropout  color (if the redshift is at the high end of the distribution
with Ly$\alpha$ in the redder filter) or it can dilute the color (if
the redshift is at  the low end with Ly$\alpha$ in the bluer filter).
In principle, this  could lead to Ly$\alpha$ being preferentially
recovered at higher redshifts.  But we minimize these biases by
simultaneously targeting the B, V, and  $i'$-dropout population.  For
example, sources with very strong Ly$\alpha$ emission at $5.5<z<5.7$
(such that the line falls in the $i'_{775}$-band) may  be scattered
out of the $i'$-drop selection but would instead appear in  V-drop
selections.  Thus, in this case, by conducting spectroscopy of
V-drops, we can account for this diffusion of Ly$\alpha$
sources. Similarly,  our B-drop sample will contain the small number
of sources at $4.5<z<4.8$  with very high strong Ly$\alpha$
contaminating the $V_{606}$-band filter (which  would otherwise have
little flux).  While we don't target U-drops, Ly$\alpha$  emission
from galaxies at $3.5<z<3.8$ does not contaminate the $B_{435}$-band
filter, so our B-drop sample should not have redshift-dependent
biases.

The redshift dependence of $x_{Ly\alpha}$ is affected not only by
evolution in the internal properties of galaxies but also by the
increase in the density of the IGM with redshift.  At $z\simeq 6$, the
IGM provides a significantly greater optical  depth to Ly$\alpha$
photons than that at $z\simeq 4$, resulting in a second order affect
on the Ly$\alpha$ fraction.   In the absence of IGM density evolution,
we would thus expect the redshift evolution  of the Ly$\alpha$
fraction to be slightly greater than derived above.  We can attempt to
estimate the variation in $x_{Ly\alpha}$ that is intrinsic to {\it
  galaxy evolution} (e.g. dust, ISM kinematics) by subtracting the
differential evolution expected from changes in IGM density.
Deconvolving  the effects of the IGM on Ly$\alpha$ radiative transfer
requires  careful modeling of the local density, velocity, and
ionisation  field (e.g.
\citealt{Santos04a,Dijkstra07a,Dijkstra07b,Zheng10}).  We delay such a
treatment to subsequent works, and instead  we follow a very simple
approach adopted in \citet{Ouchi08} which yields a very rough estimate
on the intrinsic redshift evolution of the Ly$\alpha$  fraction.  We
compute the percentage of photons absorbed by the IGM  assuming that
the blue side of  the Ly$\alpha$ line is attenuated by
$\rm{exp[\tau_\alpha(z)]}$, where  $\tau_\alpha(z)$ is the optical
depth for Ly$\alpha$ photons as  computed in \citet{Meiksin06}.   With
this approach, we find that the IGM absorbs 28, 42, and 49\% of the
Ly$\alpha$  line flux at $z\simeq 4$, 5, and 6.  In a more
sophisticated and realistic treatment, the density and ionising
background surrounding Ly$\alpha$  emitters is likely to be greater
than the mean, and infalling gas would also erase a fraction of the
Ly$\alpha$ line redward of rest-frame 1216~\AA; the combination  of
these effects can cause the redshift evolution in the transmission of
Ly$\alpha$ photons through a reionised IGM to be considerably
different than implied by our model above (e.g., \citealt{Santos04a,
  Dijkstra07b}).  We will model this effect in greater detail in the
future. For the sake of clarity, here we define the intrinsic
W$_{Ly\alpha,0,\rm{int}}$ as the rest-frame equivalent  width that
would have been observed if not for IGM attenuation, where the IGM
absorption  is taken to follow the numbers derived above.  Adopting  a
fixed intrinsic W$_{Ly\alpha,0,\rm{int}}$, we derive Ly$\alpha$
fractions as above and find that the differential redshift evolution
in $x_{Ly\alpha}$ increases to $\rm{dx_{Ly\alpha}/dz \simeq 0.12}$.
In the next section, we attempt to understand the factors likely to be
creating this redshift trend and the luminosity trend presented in the
previous section.

\subsection{The factors governing the Ly$\alpha$ escape fraction}

Earlier we demonstrated that prevalence of strong Ly$\alpha$ emission
increases toward lower luminosities and higher redshifts.  Here  we
discuss the factors that are likely governing the observed trends
prior to exploring the use of higher redshift galaxies and their line
emission as a probe of cosmic reionisation.  In \S4.3.1, we examine
whether trends in dust obscuration could potentially drive the
observed  Ly$\alpha$ fraction relations.  In \S4.3.2 and \S4.3.3, we
discuss  how the geometric distribution and kinematics of the
surrounding ISM may  impact the Ly$\alpha$ fraction trends we observe
and discuss future  spectroscopic work in progress to provide further
insight into the ISM of  feeble high-$z$ galaxies.  Finally in \S4.3.4
and \S4.3.5, we close our  discussion with a variety of factors
possibly affecting Ly$\alpha$ that  are more difficult to directly
constrain observationally.

\subsubsection{Impact of dust extinction}

\begin{figure}
\includegraphics[width=0.47\textwidth]{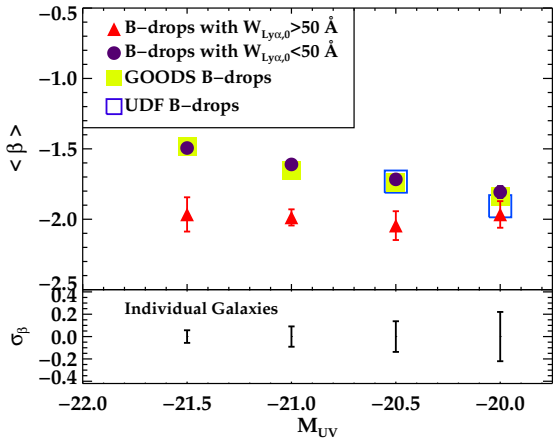}
\caption{Average UV slopes of B-drops with and without strong
  Ly$\alpha$  emission.  Average UV slopes of B-drops (grouped in bins
  of M$_{UV}$) from the spectroscopic sample with strong Ly$\alpha$
  emission (W$_{\rm{Ly\alpha}}> 50$~\AA) are denoted by red triangles,
  while the average UV slopes rest  of the B-drop sample is denoted by
  purple circles.  For reference, we overplot the average  UV slope as
  a function of M$_{UV}$ for the entire photometric sample  of B-drops
  in GOODS-N and GOODS-S from \citet{Stark09} as filled  yellow
  squares.  The average UV slope from B-drops in the Hubble  UDF is
  shown as open blue squares.  Typical uncertainties  in the UV slopes
  of individual galaxies are shown as error bars on the bottom  of the
  plot; uncertainties  in the mean take both this error and the number
  of sources in each bin into  account. }
\label{fig:dust}
\end{figure}

Previous observations have demonstrated that, among luminous LBGs,
those  objects showing Ly$\alpha$ in emission tend to display bluer UV
continuum  slopes than those with Ly$\alpha$ absorption
\citep{Shapley01,Shapley03,Vanzella09,Pentericci09,Kornei09}. While
the  presence of dust may enhance Ly$\alpha$ relative to the continuum
\citep{Neufeld91,Hansen06,Finkelstein08}, the results in Figure
\ref{fig:dust} suggest that most often, the presence of significant
quantities of dust generally leads to increased absorption of Ly$\alpha$
photons relative to the continuum.

Here, we examine whether similar trends are seen at lower
luminosities.   In order not to bias the UV colors, we determine the
slopes using filters that are not contaminated with Ly$\alpha$
emission or  IGM absorption.  Obtaining accurate  UV slopes requires
very accurate color measurements, thus we choose to focus  on the
B-drops, as the UV colors can be determined entirely from  deep $i'$
and $z$-band imaging  with ACS.  We translate the UV colors into UV
slopes using  the relation presented in \citet{Bouwens09b}:
$\beta$=5.30($i_{775}-z_{850}$) - 2.04.   We have verified that this
relation holds for a variety  of star formation histories and age
combinations that are appropriate for the $z\simeq 4-6$ population.

Given the relationship between UV color and $\beta$, it is important
to note that even a small photometric colour error translates  into
substantial uncertainty in the derived UV slope.  We thus estimate the
luminosity limit  at which the GOODS data become unreliable for
estimating UV slopes by  comparing the average UV slopes of B-drops in
GOODS with UV slope  measurements from higher S/N data in the UDF,
using the photometric catalogs of \citet{Coe06} for our UDF sample.
These results indicate that UV slopes measured from the GOODS dataset
differ significantly from those determined from the high S/N UDF
sample for B-drops with luminosities fainter than M$_{UV}\simeq
-20.5$.  Therefore, we cannot derive robust UV slopes for sources in
our GOODS spectroscopic sample that are fainter than this limit.

Concentrating on the brighter subset of objects, we examine the UV
slopes of our B-drop spectroscopic sample as a  function of UV
luminosity (Figure~\ref{fig:dust}).   The data show that galaxies with
strong Ly$\alpha$ in emission ($W_{\rm{Ly\alpha,0}}>50$~\AA) are bluer
than those systems without strong Ly$\alpha$ emission, and are
generally fit with $\beta=-2.0$ across the entire luminosity range
covered.  Following \citet{Meurer99}, this value is consistent with
very little dust obscuration  (A$_{1600}\simeq 0.5$), assuming a
Calzetti extinction curve and normal stellar populations.  The UV
slopes of the overall population of B-drops are significantly redder
than the Ly$\alpha$ emitters, but when viewed as a function of
luminosity, they grow steadily bluer  toward low  luminosities,
ranging from $\beta \simeq -1.5$ (at M$_{\rm{UV}}\simeq -21.5$) to
$\beta \simeq -1.7$ (at M$_{\rm{UV}}\simeq -20.5$).  The correlation
between UV slope and emerging UV luminosity at these redshifts was
first demonstrated in a large photometric sample in
\citet{Bouwens09b}; here we  confirm this trend with a spectroscopic
sample of dropouts.  As argued  in \citet{Bouwens09b}, the trend most
likely arises as a result of lower luminosity galaxies having less
dust obscuration.  Given that strong Ly$\alpha$ emitters  tend to
arise in galaxies with little  dust extinction, it is perhaps no
surprise that we see a larger Ly$\alpha$ fraction in low luminosity
galaxies.  Similar reasoning can also explain  the observed
redshift-dependence of the Ly$\alpha$ fraction, as UV  slopes are
found to grow steadily bluer with redshift over $3\lsim z\lsim 6$
\citep{Bouwens09b}.

If we attribute the luminosity-dependence of the Ly$\alpha$ fraction
to the variation in dust obscuration, we can derive a relationship
between mean UV slope and Ly$\alpha$ fraction.  From Figure
~\ref{fig:dust}, we find that $d\beta/dM_{\rm{UV}}\simeq -0.23$ over
$-21.5<M_{\rm{UV}}<-20.0$.  This is similar, but slightly steeper,
than the value derived in \citet{Bouwens09b}.  Over this same
luminosity range, we derive $dx_{Ly\alpha}/dM_{\rm{UV}}=0.29$.
Assuming that dust extinction drives the  evolution in the Ly$\alpha$
fraction, we derive $dx_{Ly\alpha}/d\beta=-1.2$.   While this result
was derived from the luminosity-dependence of  $x_{Ly\alpha}$, it
should also apply to the redshift evolution if dust obscuration were
to dominate the redshift-dependent  trends.  Measurements of similarly
luminous $z\simeq 6$  LBGs \citep{Bouwens09b} suggest that the average
UV slopes grow bluer by $\Delta\beta \simeq -0.6$ from $z\simeq 4$ to
$z\simeq 6$.   If dust evolution dominates the redshift-dependence of
the Ly$\alpha$ fraction and the evolution in UV slopes, the above
relationship suggests that $x_{Ly\alpha}$ should increase by $\Delta
x_{Ly\alpha} = 0.84$.  The fact that the actual redshift evolution is
significantly less rapid ($\Delta x_{Ly\alpha} = 0.24$ after roughly
accounting for IGM absorption) can be explained by several factors.
For example,  it is likely that the luminosity dependence of
$x_{Ly\alpha}$ is driven by more than just dust obscuration (see the
following sections for a discussion).  Additionally as UV slopes grow
bluer than $\beta=-2.0$, the variation in the UV slope is  possibly
driven by factors other than dust obscuration (e.g.,
\citealt{Stanway05, Bouwens10b}).  Secondly, given that $x_{Ly\alpha}$
cannot be larger than 1.0, the  differential growth must slow down as
$x_{Ly\alpha}$ increases; hence any extrapolation of the
$x_{Ly\alpha}$ relations below our  completeness limits is highly
uncertain.  Finally, as emphasized in \S4.2,  our simple treatment of
IGM absorption may incorrectly estimate the evolution in the
Ly$\alpha$ fraction that is intrinsic to galaxy processes. 

In summary, these results indicate that the variation in dust
extinction with luminosity and redshift likely plays an important role
in governing the observed evolution and luminosity-dependence in the
Ly$\alpha$ fractions.  But, not surprisingly, it seems likely that
additional factors play a role  in governing $x_{Ly\alpha}$ as well.
We discuss these in more detail  below.

\subsubsection{The hydrogen covering fraction}

After Ly$\alpha$ photons escape the H II regions where they were
created, they must diffuse through neutral gas and dust at larger
radii.  This could include gas that is participating in outflows
surrounding the galaxy in addition to gas that is being accreted onto
the galaxies.  Rest-UV spectra of high-redshift star-forming galaxies
reveal low-ionisation absorption lines that are generally blue-shifted
by $\Delta v \simeq -200$ km s$^{-1}$ \citep{Shapley03, Steidel10}
relative to the centre of rest, indicating that outflowing neutral gas
is nearly always present, while accreting gas appears less prevalent.
The geometrical  distribution and kinematics of this outflowing
material plays a  crucial role in governing the escape of Ly$\alpha$
photons.  This is clearly evidenced by the fact that Ly$\alpha$
photons  are typically observed to  be redshifted by $\simeq 400-500$
km s$^{-1}$  with respect to the centre of rest \citep{Steidel10}.
This result  implies that the majority of Ly$\alpha$  photons to
escape through the absorbing material along the  line of sight are
those that achieve a sufficient (redshifted) velocity such that they
can travel through the intervening material without resonantly
scattering.  This is most easily accomplished in a model in which  the
Ly$\alpha$ photons that are observed are those that have been
``backscattered'' off of the ouflowing material on the far side of the
galaxy (e.g., \citealt{Shapley03,Steidel10}).  Hence, it is likely
that the increased prevalence of Ly$\alpha$ emitters among low
luminosity  galaxies  tells us something about the distribution and/or
kinematics of the  ISM of feeble sources.  In this  section, we
discuss the relationship of the ISM distribution, dust obscuration,
and Ly$\alpha$ emission in luminous LBGs as found in previous work,
and then consider whether a similar picture is likely to hold for low
luminosity galaxies; in the following section, we  focus on the
kinematics of the ISM.  

Even for luminous sources, observational constraints on the
distribution of the blueshifted neutral absorbing gas in the immediate
vicinity of high redshift galaxies are limited.  The spectroscopic
study of $z\simeq 3$ LBGs conducted in \citet{Shapley03} revealed a
strong correlation between the equivalent width of low-ionisation
interstellar absorption lines (W$_{\rm{LIS}}$) and that of Ly$\alpha$,
in  the sense that the strongest  Ly$\alpha$ emitters tend to have the
least absorption by the low-ionisation interstellar medium.  As the
absorption lines are highly saturated, \citet{Shapley03} note that the
trend in W$_{\rm{LIS}}$  is due either to variations in the velocity
width or the covering fraction of absorbing gas, with the latter
argued to be the dominant factor, such that sources showing strong
Ly$\alpha$ emission, on average, have the patchiest distribution of
absorbing gas covering the continuum source (at least along the  line
of sight).   Galaxies also show a strong correlation between
W$_{\rm{LIS}}$ and E(B-V), which \citet{Shapley03} argue implies a
significant fraction of the dust which reddens the stellar continuum
(and absorbs Ly$\alpha$ photons) is located {\it within} the
outflowing neutral gas.  As noted in \citet{Steidel10}, from a
theoretical standpoint, the exact effect  of a non-uniform covering
fraction on the observed Ly$\alpha$ flux is not obvious.  But
regardless, these observations thus suggest a scenario in which strong
Ly$\alpha$ emission is generally coupled with low dust extinction and
a low hydrogen covering fraction, a picture supported by  recent
observations of strongly-lensed LBGs at $z\simeq 3$
\citep{Quider09,Quider10} for which direct measures of the covering
fractions  of various ions are available.

If similar trends are present in low luminosity galaxies, then the
fact that   Ly$\alpha$ is much more common in feeble galaxies may
imply that these sources  typically have lower covering fractions of
absorbing gas than more luminous LBGs.  Addressing whether this is
indeed the case requires deep spectroscopy of UV-faint systems, and
thus is perhaps only feasible via studies of gravitationally-lensed
galaxies.  Some progress can be made with current field samples
however. In particular,  we can determine whether  the coupling
between strong Ly$\alpha$ emitters and weak ISM absorption is also
present for feeble galaxies by creating composite spectra of
Ly$\alpha$  emitters binned by $\rm{M_{\rm{UV}}}$.  We will present
the results of this analysis in a subsequent  paper (Stark et
al. 2010, in preparation).  Some indication that the correlation is in
place at low intrinsic luminosities is already apparent from the
$i'$-drop composite spectra presented in \citet{Vanzella09}.  As all
the $i'$-drops in their sample are UV faint ($<\rm{M_{\rm{UV}}>\simeq
  -20}$), this spectrum provides  insight into the properties of
feeble sources.  Owing to the faintness of  the $i'$-drops, the
composite is dominated by strong Ly$\alpha$ emitters  and shows very
weak interstellar absorption lines.  Higher S/N spectra are required
to ensure that the absorption lines are saturated and to quantify the
absorption line equivalent widths.  

Finally, while admittedly speculative, we note briefly that if a low
hydrogen covering fraction is more common for low  luminosity
galaxies, it may also enable Lyman continuum  photons to more easily
escape from feeble systems.  Naively,  this statement seems
contradictory since the presence of Ly$\alpha$ photons  stems from the
absorption of ionising photons.  However high-redshift galaxies  are
clearly not perfect H II regions. In practice, a significant fraction
of ionising photons are absorbed in the ionised regions surrounding
the massive stars, leading to the production of Ly$\alpha$ photons.
Both Ly$\alpha$ photons as well as any escaping ionising photons will
then approach the surrounding neutral ISM, much of which is likely
moving at great speeds with respect to the stars.  Clearly, those
systems with   significant holes in their surrounding distribution of
hydrogen will leak  a larger fraction of ionising radiation.   Indeed,
the combination of strong Ly$\alpha$ and significant ionising  photon
escape fractions are seen in observations at $z\simeq 2$
\citep{Shapley06}.  Intriguingly, recent results reveal a possible
trend toward greater Lyman  continuum leakage in sources with low UV
continuum  luminosities (Steidel et al. 2010, in preparation).   While
much work is still required to verify the luminosity-dependence in the
hydrogen covering fraction and  the Lyman continuum escape fraction,
these results strongly motivate detailed  study of the physical
properties of  low luminosity galaxies.

\subsubsection{Kinematics of ISM}

As we discussed above, Ly$\alpha$  is affected not only by the
geometry and column density of the  dust and hydrogen around it but
also by the velocity distribution of the surrounding neutral HI.
Recently, \citet{Steidel10} has compared the kinematics of the
absorbing  gas (extracted from the properties of rest-UV absorption
lines) and  the properties of Ly$\alpha$ as a function of total baryon
mass for a sample of $z\simeq 2$ UV continuum  selected galaxies.
These results  reveal that the low mass subset shows significantly
stronger Ly$\alpha$  emission, as we would expect from Figure
\ref{fig:lya_frac}, on the assumption  that the feeble sources that we
have studied spectroscopically are generally  lower mass systems.
\citet{Steidel10} demonstrate that the absorbing medium is
significantly  different for these two mass subsets, with the more
massive galaxies  more often having a significant component of
interstellar absorption at  velocities close to the systemic redshift
(and often extending to positive  velocities).  These high mass
systems not only have weaker Ly$\alpha$ emission but  the redshifted
velocity of Ly$\alpha$ is greater than in low mass galaxies.  The
nature of this excess absorption component at zero velocity for
massive  galaxies is unclear; as discussed in detail in
\citet{Steidel10} it could possibly be outflowing material which has
stalled due perhaps to the larger  gravitational potential, or
alternatively infalling gas. 

Regardless, it is clear that there are significant mass-dependent
variations in the kinematics of the absorbing gas, particularly at
$v=0$, such that lower mass galaxies transmit a larger fraction of
their Ly$\alpha$ photons to  earth.  These trends, together with those
seen  in dust obscuration and perhaps hydrogen covering fraction,  may
play a large role in governing the luminosity-dependence of the
Ly$\alpha$ fraction seen in Figure \ref{fig:lya_frac}.  Future
spectroscopic  work of feeble galaxies which are magnified via
gravitational lensing can  help confirm these mass-dependent trends in
the ISM kinematics.

\subsubsection{Addtional Factors Governing the Ly$\alpha$ fraction}

We conclude by mentioning several additional factors which may
contribute  to the observed Ly$\alpha$ fraction trends.  Firstly, the
Ly$\alpha$ equivalent width is highest in the earliest  stages of star
formation \citep{Charlot93,Leitherer99}, when the  contribution from
massive, young stars is at its greatest.  Studies  of UV faint
narrowband LAEs have generally shown that these sources  have younger
ages and higher specific star formation rates than  more luminous LBGs
(e.g. \citealt{Pirzkal07,Ono10}).  However, a  recent study of
spectroscopically-confirmed LBGs has demonstrated that  those sources
with strong Ly$\alpha$ emission tend to be older than those  without
Ly$\alpha$ emission at similar UV luminosities
\citep{Shapley01,Kornei09}, leading the authors to suggest a  physical
picture whereby Ly$\alpha$ emission escapes more easily once large-scale
outflows have had time to reduce the dust covering fraction
sufficiently.  The sample presented in this paper allows us to extend
the work of \citet{Kornei09} to lower continuum  luminosities, which
we plan to present a future study (Stark et al. 2010,  in
preparation), taking care to  account for the possible contribution of
nebular emission lines to the mid-IR  flux
(e.g. \citealt{Schaerer09}).

Additionally, Ly$\alpha$ equivalent width becomes larger at low
metallicities  owing to the increased ionising flux and harder UV
spectra (e.g., \citealt{Schaerer03}).  Thus if low luminosity galaxies
have much lower metallicities than luminous  systems, then we may
expect to see an increased fraction of strong line  emitters among UV
faint systems.  Only via future direct measurements of the metallicity
of low luminosity galaxies at these redshifts will we be able to
predict the magnitude of this effect on the luminosity dependence of
the  Ly$\alpha$ fraction.

Finally, it is, in principle, conceivable that the trends could be
driven by variation in the stellar IMF with luminosity and redshift.
As has been demonstrated in previous studies \citep{Schaerer03},
top-heavy IMFs can boost the equivalent widths of Ly$\alpha$ relative
to that expected for normal stellar populations.  While several
studies  have argued that the IMF may vary with redshift
(e.g. \citealt{Dave08}), there is little direct evidence of such a
variation.  

\subsection{Comparison with narrowband Ly$\alpha$ emitter studies}

Finally, we compare our results to those determined from studies of
narrowband-selected Ly$\alpha$ emitters (LAEs) in the same redshift
range \citep[e.g.][]{Shimasaku06,Kashikawa06, Ouchi08}.  We first
contrast the two samples and examine whether the large fraction of
strong Ly$\alpha$ emission seen among our UV faint dropouts is
consistent with the independently-determined luminosity functions of
LAEs.  Then we consider whether we can use the Ly$\alpha$ trends
discovered in our UV continuum samples to explain the lack of redshift
evolution in the luminosity functions of the LAE population, which
contrasts markedly with the strong evolution seen in the LBG samples.

It is commonly asserted that narrowband LAE samples are fainter than
UV continuum LBG samples.  While this is true if one compares the LAE
samples to the well-studied spectroscopic $z\simeq 3$ population
\citep[e.g.][]{Shapley01,Shapley03}, current photometric LBG samples
extend to much fainter UV luminosities (e.g., M$_{UV}\simeq -16$ at
$z\simeq 4$  in the Ultra Deep Field), and as we have discussed above,
our spectroscopic observations take advantage of these faint samples,
extending to $M_{\rm{UV}}\simeq -18$ (Figure~2).  Many LAEs identified
in typical ground-based surveys are below the UV continuum magnitude
limits  and require stacking to determine the typical continuum
luminosity.  In  a recent analysis of a large sample of $z=3.1$ LAEs,
\citet{Ono10} have  shown that the stacked continuum magnitude is
$i'\simeq 27$, corresponding  to $M_{\rm{UV}}\simeq -18.6$, comparable
to the UV continuum luminosity of the  faintest galaxies in our
spectroscopic sample.  Thus by probing down the LBG  luminosity
function in our DEIMOS survey, we are able to directly compare  the
Ly$\alpha$ trends discovered in both populations.

Consider the percentage of LBGs at a given redshift that show strong
Ly$\alpha$ emission.  Assuming that LBGs form the parent population of
Ly$\alpha$ emitters, we can compute the expected number  density of
strong Ly$\alpha$ emitters as a function of M$_{\rm{UV}}$.  Our
spectroscopic data suggests that 10\% of B-drops with M$_{UV}\simeq
-21$  show Ly$\alpha$ emission with $W_{Ly\alpha,0}>50$~\AA.  Based on
the number  density of LBGs in this luminosity range, these data
predict that the density  of strong line emitters with
$M_{\rm{UV}}\simeq -21$ should be $\simeq 1.8\times$10$^{-5}$
Mpc$^{-3}$ mag$^{-1}$.  This value is very similar to the abundance of
LAEs in this luminosity range from the UV luminosity functions  of
narrowband Ly$\alpha$ emitters at $z\simeq 4$ (1.2$\times$10$^{-5}$
Mpc$^{-3}$) \citep{Ouchi08}.  Clearly this calculation is very
uncertain owing to  the different selection and equivalent width
limits, and is meant as a zeroeth order comparison.  We note that of
the 26  $z=3.7$ LAEs in \citet{Ouchi09}, 23 have rest-frame equivalent
widths  greater than our threshold of $\simeq 50$~\AA, so the LAE
sample  isn't probing significantly further down the equivalent width
distribution  then our LBG sample.  At fainter continuum luminosities
($M_{\rm{UV}}>-20$), the agreement begins to break down, as our
measured  Ly$\alpha$ fractions predict a larger abundance of LAEs than
inferred from the \citet{Ouchi08} UV luminosity functions. This is not
surprising given that the \citet{Ouchi08} is only computed over
$-21.7<M_{\rm{UV}}<-19.7$, (where the data are sufficiently complete),
requiring M$^{\star}_{UV}$ and  $\alpha$ to be fixed.  

Finally we consider whether the observed evolution of  the Ly$\alpha$
population over $3\lsim z\lsim 6$ is qualitatively consistent with the
trends suggested by the luminosity and redshift-dependence of the
Ly$\alpha$ fraction.  The rapid increase in the Ly$\alpha$ fraction
toward lower luminosities suggests that deep narrowband Ly$\alpha$
samples will  generally be dominated by UV faint galaxies.  Since the
redshift evolution of the number density of UV faint LBGs is much less
rapid than that of luminous LBGs, we expect the number density of
Ly$\alpha$ emitters (which should be weighted more toward UV faint
sources)  to decrease less rapidly than LBG samples.  This trend is
enhanced by the increase in the prevalence of Ly$\alpha$ emission in
LBGs over this redshift interval.  Hence given the luminosity and
redshift-dependent trends in the Ly$\alpha$ fraction, it is not
necessarily surprising that \citet{Ouchi08} reveal that the observed
Ly$\alpha$  luminosity function does not evolve significantly with
redshift over $3.1<z<5.7$.   

\section{Implications for Reionisation} 

In the previous section, we showed that strong  Ly$\alpha$-emitters
become more common between $z\simeq 3$ and $z\simeq 6$.  We argued
that this trend is  likely driven in part by a decrease in dust
extinction and the covering fraction of hydrogen surrounding the H II
regions where Ly$\alpha$ photons are  originally produced.  The first
estimates of UV slopes at $z\gsim 6$ imply that the dust obscuration
continues  to decline to $z\simeq 7-8$.  Given these trends, the
expected signal of reionisation (decreasing the prevalence of
Ly$\alpha$ emitters) should be readily apparent in deep spectroscopy
of newly discovered $z_{850}$-band and $Y_{105}$-band dropouts
\citep[e.g.][]{McLure10,Oesch10,Bunker10,Bouwens10b,
  Wilkins10,Wilkins10b}.

Given the increase we find in the Ly$\alpha$-emitting fraction over
$3\lsim z\lsim 6$ (Figure~\ref{fig:lya_frac})  it is now interesting
to examine whether there is any decline seen beyond $z\simeq 6$ such
as might arise from an increasing neutral fraction in the IGM.
Although our first set of $z$-drop observations were conducted in poor
conditions, we can still provide preliminary constraints on
$x_{Ly\alpha}$ if we adopt a sufficiently bright equivalent width and
magnitude limit.  The flux limits from the October 2009 run indicate
that detection requires an equivalent width of at least 75~\AA~
(rest-frame) for sources in the luminosity bin $-20.5 \lsim
M_{\rm{UV}}\lsim -19.5$ (\S2.1).  The Ly$\alpha$ equivalent widths of
sources fainter  than this limit (all but two of the $z$-drops in the
UDF) are not usefully constrained.  Also, owing to the increased noise
stemming from poor atmospheric conditions it is not possible to detect
Ly$\alpha$ with typical equivalent widths in any of the sources
studied at redshifts beyond $z\simeq 6.65$.  Over the redshift  range
in which we are sensitive to Ly$\alpha$ emission, completeness
simulations indicate that recovery rate of lines with
$W_{Ly\alpha,0}>75$~\AA~ is $\simeq 50$\%.  The photometric redshifts
derived from the broadband SEDs  indicate that five of the eight
$z$-drops are likely to lie at redshifts above  $z\simeq 6.65$. If
this is true, these objects would escape detection in our DEIMOS
spectra.  We take this possibility into account in our discussion of
the Ly$\alpha$ fractions below.  

\begin{figure}
\includegraphics[width=0.47\textwidth]{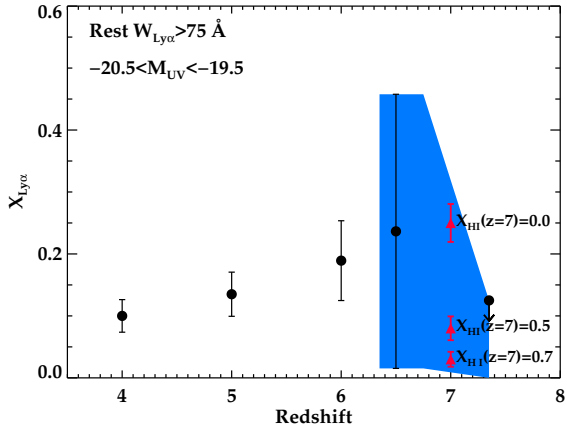}
\caption{Fraction of spectroscopic dropout sample with
  -$\rm{20.5<M_{\rm{UV}}<-19.5}$ showing strong Ly$\alpha$ emission
  (W$_{\rm{Ly\alpha}} > 75$~\AA) over $4<z<7$. We estimate the
  Ly$\alpha$  fraction at $z\simeq 6.5$ from a recent DEIMOS run
  targeting WFC3 $z$-drops.   Owing to poor atmospheric conditions on
  this run, statistics are insufficient  to robustly constrain the
  evolution in $x_{Ly\alpha}$.  However  in contrast, the implied
  Ly$\alpha$ fraction from the lack of Ly$\alpha$  detection in a
  sample of 8 lensed $z$-drops studied with near-IR spectroscopy in
  \citet{Richard08} and \citet{Kneib04} is shown by the upper limit at
  $z=7.3$. If these candidates are real and at high-$z$, then  the
  implied Ly$\alpha$ fraction at $z\simeq 7$ is lower than measured at
  $z\simeq 6$.  Red triangles and associated errors correspond to
  measurements  that will be possible if a large percentage of
  $z$-drops expected to be discovered with future observations
  (e.g. ULTRA-VISTA and WFC3 MCT) are  followed up with spectroscopy.
  The quoted neutral fractions correspond to those implied from a
  mapping  derived from \citet{McQuinn07} and are only  valid at
  $z\simeq 7$.}
\label{fig:lya_zev_z47}
\end{figure}

Figure \ref{fig:lya_zev_z47} shows the overall evolution in the
Ly$\alpha$ fraction with  redshift within this restricted luminosity
range.  The fraction of LBGs with Ly$\alpha$ emission above our chosen
threshold grows steadily over $4\lsim z\lsim 6$, reaching nearly  20\%
at $z\simeq 6$.   Based on our discussion in the previous section,  we
argue that this net change likely arises from a combination of the
redshift evolution in the dust and hydrogen covering fractions (which
increases the Ly$\alpha$ fraction toward higher redshift)  and the IGM
density (which decreases the Ly$\alpha$ fraction toward higher
redshift).  

If the line shown in Figure~\ref{fig:zdrop_spec} is Ly$\alpha$, and if
all $z$-drops in our restricted luminosity range
($\rm{-20.5<M_{\rm{UV}}<-19.5}$) lie at $z\lsim 6.65$ (as is necessary
for detection with DEIMOS in the October dataset), then our
measurements would indicate that the fraction from the October run
with $\rm{W_{Ly\alpha}>75}$~\AA~ (corrected for incompleteness) is
marginally greater than that measured at $z\simeq 6$, consistent with
the slow increase over $4<z<6$.   However, given the very large errors
in the Ly$\alpha$ fraction, the  $z\simeq 6.5$ measurements are also
fully consistent with a  significant decline over $6.0<z<6.5$.
However  we point out that the measured Ly$\alpha$ fraction may
actually be larger than we estimate here, for if five of the $z$-drops
are indeed at $z>6.65$, as suggested by their photometric redshifts,
then the Ly$\alpha$ fraction at $z\simeq 6.5$ would be $\simeq
2.6\times$ larger than the datapoint in Figure \ref{fig:lya_zev_z47}
indicates.  Clearly additional spectroscopy in improved atmospheric
conditions is required for robust statements.

Beyond $z\simeq 6.6$, existing datasets imply that the Ly$\alpha$
fraction may begin to decline.  In contrast to the abundant samples of
robust LBGs now  in place at $z\simeq 7$ and beyond
\citep[e.g.][]{McLure10,Bunker10,Oesch10,Bouwens10b}, searches for
Ly$\alpha$ emitters at $z>7$ with narrowband filters have yet to
reveal  a large population of line emitters.  While candidates have
been identified (e.g. \citealt{Stark07b,Hibon09}), confirmation has proven
challenging.  Indeed, the  highest redshift Ly$\alpha$ emitter with
robust spectroscopic verification lies at $z=6.96$ \citep{Iye06}.
Updated data from this survey (Ota et al. 2008) suggest that the
number density of Ly$\alpha$ emitters at $z=7.0$  is only 17\% of that
at $z=6.6$.   Attributing this decline in number density to a decrease
in Ly$\alpha$ transmission  (ignoring the contribution from the
declining galaxy number density),  we would expect see strong
differential evolution in the Ly$\alpha$ fraction over $6.6\lsim
z\lsim 7.0$.  In principle, the DEIMOS observations should have easily
detected such a decline in the initial sample of 17 $z$-drops observed
in October 2009. Unfortunately, as discussed earlier,  the poor
conditions prohibit detecting Ly$\alpha$ at $z\gsim 6.6$ for  the
faint sources considered in this work.  Therefore we are unable  to
verify this result with our current DEIMOS dataset. 

However, the only detailed spectroscopic survey of $z$-drops currently
in the  literature \citep{Richard08} does offer tentative support for
the downward  trend in the Ly$\alpha$ fraction.  In their paper, seven
gravitationally-lensed $z$-drops were  observed with the NIRSPEC
spectrograph on Keck \citep{Mclean98}, and none showed evidence  for
Ly$\alpha$ emission (with rest-frame equivalent width limits as low as
$\simeq 30$~\AA).  Likewise, a similar absence of Ly$\alpha$ was seen
in the $z\simeq 6.8$ gravitationally  lensed galaxy in Abell 2218
\citep{Kneib04, Egami05}.  The typical luminosities of these sources
are  considerably lower than the restricted range used in Figure
\ref{fig:lya_zev_z47},  hence one would expect them to have Ly$\alpha$
fractions that are significantly greater than those observed at lower
redshifts in this figure.  In contrast, we measure an upper limit to
the $z\simeq 7$ Ly$\alpha$ fraction  which is suggestive of a
significant decline over $6<z<7$.  At the very large equivalent width
thresholds adopted in  this figure, it is unlikely that this decline
could be  produced by incompleteness owing to the presence of OH
emission lines.  While these NICMOS dropouts are generally not as
robust as the $z$-drops being identified with WFC3 (see
\citealt{Bouwens09a}), the lack of Ly$\alpha$ is  tantalising in light
of the narrowband LAE results \citep{Ota08}.  Such a marked decline in
the number density of Ly$\alpha$ emitters should easily  be detected
in future follow-up of current WFC3 $z$-drop samples.

We now  explore the potential ramifications that these results may
have for reionisation and discuss the possibilty of improving and
extending the $x_{Ly\alpha}$ measurements in the future.   We first
discuss how one may use these measurements to quantify  changes in the
IGM.   Various theoretical studies have examined how an increase in
the neutral  fraction of the IGM impacts the transmission of
Ly$\alpha$ photons (e.g.,
\citealt{Santos04a,Furlanetto06,McQuinn07,Mesinger08}).  In this
paper, we consider the models of \citet{McQuinn07}, which  make use of
200 Mpc radiative transfer simulations to compute the effect of
inhomogenous reionisation on the transmission of Ly$\alpha$ photons
and the clustering of Ly$\alpha$ emitting galaxies.  In Figure 4 of
their paper, \citet{McQuinn07} present the Ly$\alpha$ luminosity
function at $z=6.6$ for various HI fractions, ranging from
$x_{HI}=0.00$ to $x_{HI}=0.88$.  The greater the HI fraction, the more
the amplitude of the  Ly$\alpha$ luminosity function decreases with
respect to the fully-ionised case. The relative amplitude of the
luminosity functions for various ionisation fractions is directly
related to the decreased transmission of Ly$\alpha$  photons
associated with reionisation of the IGM.  For example, the  amplitude
of the Ly$\alpha$ luminosity function decreases by a factor of $\simeq
120$  when the HI fraction increases to 0.88.  

Our goal is to translate the measured Ly$\alpha$ fractions into
estimates of the ionisation state of the IGM.  We can convert the
\citet{McQuinn07} results into a mapping between Ly$\alpha$ fraction
and x$_{HI}$ as follows.  First, we naively assume that in the absence
of any change in the ionisation of the IGM, the Ly$\alpha$ fraction
would not evolve between $z\simeq 6$ and $z\simeq 7$ (see below  for
an alternative approach).  We then assume that the predicted decline
in the amplitude of the Ly$\alpha$  luminosity function (with respect
to when the IGM is fully ionised)  will result in an identical
decrease in the Ly$\alpha$ fraction.  Hence  if the HI 
fraction is 0.88 at $z\simeq 7$, we should expect the  Ly$\alpha$
fraction to be 120$\times$ lower than its measured value at  $z\simeq
6$, when the IGM is ionised.  With these assumptions, the current
measurements  at $z\simeq 6.5$ indicate that the IGM is still largely
ionised.  

However given that the Ly$\alpha$ fraction evolves with redshift  in
the absence of changes in the IGM ionisation state, our 'no-evolution'
assumption above is clearly simplistic.  Indeed if the dust
obscuration continues to evolve over $z\gsim 6-8$ (as  implied by
\citealt{Bouwens10b}), then in the absence of IGM ionisation  state
evolution, the Ly$\alpha$ fraction is likely to be even larger at
$z\simeq 6.5-7$.  If this is the case, then the mapping between
Ly$\alpha$ fraction and IGM ionisation state discussed above would be
incorrect.  To account for this, we extend the smooth evolution  seen
in the Ly$\alpha$ fraction over $4\lsim z\lsim 6$ (when the IGM is
highly ionised) to $z\simeq 6.5$ and $z\simeq 7$.  We adopt this
prediction as the baseline $x_{Ly\alpha}$ value consistent with an
ionised IGM.  We associate deviations below this value with evolution
in the IGM ionisation  state in a similar manner as above.  Following
this approach, the lack of Ly$\alpha$ in the \citet{Richard08} sample
and \citet{Kneib04} source suggests $\rm{x_{HI}>0.3}$ at $z\simeq 7$.

With new ground-based surveys (e.g. UltraVISTA) and WFC3 programs set
to reveal hundreds of $z$ and Y-drops in the next several years, it
should be possible to derive accurate Ly$\alpha$ fractions over the
redshift range $6\lsim z\lsim 8$.  Indeed, we estimate that feasible
campaigns  with optical and near-IR multi-object spectrographs should
easily be  able to detect Ly$\alpha$ equivalent widths as low as
30-50~\AA~ (rest-frame) for a sample of 200 $z$ and Y-band dropouts
with continuum magnitudes  spanning $26\lsim m\lsim 28$, enabling much
improved statistics on the fraction of LBGs with Ly$\alpha$ emission
and the possibility to measure Ly$\alpha$ fractions as low as $\simeq
1$\%.  With such a spectroscopic sample,  it will be easy to identify
whether the Ly$\alpha$ fraction at $6.5\lsim z\lsim 8$  shows a
deficit with respect to measurements at $z\simeq 5-6$.  In fact, this
dataset would enable the characterisation of Ly$\alpha$ fractions more
than 10$\times$ lower than those expected at $z\simeq 7$ in the
absence  of any change in the IGM ionisation state
(Figure~\ref{fig:lya_zev_z47}).  With these  results, it should be
trivial to identify the differential evolution in  the Ly$\alpha$
fraction that would be expected from reionisation; in the context of
the simulations of \citet{McQuinn07}, these results enable the IGM
ionisation state to be tracked between fully ionised ($x_{HI}\simeq
0.0$) and almost completely neutral ($x_{HI}\simeq 0.8$).  

\section{Summary and Conclusions}

In this paper, we have presented a new technique aimed at constraining
reionisation.  By measuring the luminosity-dependent fraction of
continuum  dropouts with strong Ly$\alpha$ emission at $z\simeq 4$, 5,
6, 7, and so on, we are sensitive to sudden changes in the
transmission of Ly$\alpha$ photons that would be expected if the
ionisation state of the IGM changes, complementary to past  efforts to
link evolution in the Ly$\alpha$ luminosity function to reionisation
\citep[e.g.][]{Malhotra04,Kashikawa06,Iye06,Ota08}.  The spectra not
only provide information on Ly$\alpha$ but also reveal constraints  on
the absorbing gas along the line of sight and provide a robust
spectroscopic sample for derivation of stellar populations
(e.g. stellar  masses, ages). Through this information, we seek to
simultaneously provide insight  into the evolution of the most
important factors governing Ly$\alpha$ transmission  (dust
obscuration, HI covering fraction and kinematics),  and to thereby
better isolate the impact of the IGM ionisation state  on the evolving
Ly$\alpha$ fraction.

To complete this Ly$\alpha$ fraction test, we have conducted a large
Keck/DEIMOS spectroscopic  survey of dropouts in the GOODS-S and
GOODS-N fields.  The current Keck/DEIMOS sample includes 268 B-drops,
95 V-drops, 64 $i'$-drops,  and 17 $z$-drops and uniquely extends to
faint continuum magnitudes (M$_{\rm{UV}}\simeq -18$).  When combined
with existing VLT/FORS2 spectroscopic observations of B, V, and
$i'$-drops  in the two GOODS fields, our sample contains 627 dropouts.
We summarize our key findings below.

1. We present tentative spectroscopic identification of a $z$-drop
identified in the GOODS ERS imaging CDF-S \citep{Wilkins10}.  The
galaxy shows a 7$\sigma$ emission feature at $z=6.44$, consistent with
the photometric redshift  expected from the broadband imaging.  If
confirmed, this  would be the first WFC3 $z$-drop with spectroscopic
identification in the literature.  Future deep spectroscopic study of
WFC3 dropouts should vastly increase the number of Ly$\alpha$ emitters
in this redshift regime.

2. The prevalence of Ly$\alpha$ emitters is greater among low
luminosity  galaxies.  We find that the fraction of dropouts with
Ly$\alpha$ equivalent  widths in excess of 50~\AA~ increases from
$\simeq 10$\% at M$_{\rm{UV}}\simeq -21.5$  to $\gsim 40$\% for
sources fainter than M$_{\rm{UV}}\simeq -19.5$.  This result  is
consistent with previous studies comparing the UV and Ly$\alpha$
luminosity function for samples of narrowband Ly$\alpha$ emitters
\citep{Ouchi08}. This finding demonstrates that a large fraction of
galaxies  at the faint-end of dropout samples are likely very similar
to the high-EW and UV-faint  LAEs found via narrowband filters.

3. We find that the fraction of strong Ly$\alpha$ emitters at fixed
luminosity moderately increases toward earlier time in the redshift
range $3\lsim z\lsim 6$.  Binning our sample in two redshift ranges
($3.0<z<4.5$) and ($4.5<z<6.0$), we find that the fraction of
Ly$\alpha$ emitters with rest-frame equivalent width in excess of
50~\AA~ increases in each luminosity bin considered.

4. We examined the possibility that these luminosity and
redshift-dependent trends are driven by evolution in the dust
obscuration of high-redshift galaxies.  We find that the $z\simeq 4$
sources  with strong Ly$\alpha$ emission have considerably bluer UV
slopes, implying significantly less  dust extinction.  This result is
consistent with a range of observational studies (e.g.,
\citealt{Shapley03,Pentericci09,Kornei09}).  In light of recent
results revealing that dust obscuration decreases toward lower
luminosities  and higher redshifts \citep{Reddy09,Bouwens09b}, it
appears that dust  evolution plays a major role in governing the
Ly$\alpha$ fraction.

5. We discuss the possibility that the covering fraction of hydrogen
may be lower in low luminosity galaxies.  In more luminous galaxies,
strong Ly$\alpha$ emission is generally coupled with low dust
extinction  and low equivalent width ISM absorption lines
\citep{Shapley03},    the latter of which appears to arise from a
non-uniform covering fraction  of neutral hydrogen \citep{Quider09}.
Given the increased prevalance of strong  Ly$\alpha$ emitters and low
dust obscuration  at low luminosity, this result may suggest that
galaxies with a non-uniform covering fraction may be much more common
among UV-faint systems.  Strong Ly$\alpha$ emitters with UV faint
continua  do appear to show weak low-ionisation ISM absorption (Stark
et al. 2010, in  preparation), but higher S/N and  higher resolution
spectra of gravitationally lensed galaxies are required to directly
measure the covering fraction.  We note that if the covering fraction
of hydrogen is indeed lower for low luminosity galaxies, this would in
turn imply that Lyman continuum photons may more easily escape from
feeble  galaxies, a trend that  appears to be seen in  $z\simeq 3$
galaxies with  deep Keck spectra (Steidel et al. 2010, in
preparation).

6. We have measured the fraction of $z$-drops with strong Ly$\alpha$
emission in our spectroscopic sample, enabling us to constrain
$x_{Ly\alpha}$  at $z\simeq 6.5$.  The estimated Ly$\alpha$ fraction
appears consistent  with the redshift trend seen over $3\lsim z\lsim
6$.  However the current sample is far too small to rule out a decline
to $z\simeq 6.5$.  Efforts to extend these  studies to $z\simeq 7$
have  thus far failed to locate Ly$\alpha$ emission.  The lack of
strong Ly$\alpha$ in the Keck spectra of 7 candidate  LBGs in Richard
et al. (2008) and the  lensed $z\simeq 6.8$ source in \citet{Kneib04}
provides tantalising evidence that the Ly$\alpha$ fraction declines by
$z\simeq 7$.   The availability of more robust WFC3/IR $z\simeq 7$
samples \citep[e.g.][]{Oesch10,McLure10,Bunker10} enables this
possibility to be readily tested with the new generation of
near-infrared multi-object spectrographs.

7.  Using recent simulations linking the evolution in the abundance
of Ly$\alpha$ emitters to the IGM ionisation state \citep{McQuinn07},
we examine the possiblity of placing constraints on reionisation with
our spectroscopic sample.  While we emphasize that the results are
strongly model-dependent, if the evolution in the Ly$\alpha$ fraction
at $z\simeq 7$ were to hold up, the \citet{McQuinn07} simulations
would  suggest  that the neutral fraction of the IGM is $x_{HI} \gsim
0.3$ at  $z\simeq 7$.  More robust  estimates of the $z\simeq 6.5-7$
Ly$\alpha$ fractions are needed to  confirm these preliminary
findings.

The recent emergence of new photometric samples of dropouts at $z\gsim
7$  with the WFC3 onboard HST provides a great deal of promise for
extending  this test in the future.  Long exposures with upcoming
near-infrared multi-object  spectrographs will enable significant
detections of strong Ly$\alpha$ emission  lines from sources as faint
as $J\simeq 28$, allowing the Ly$\alpha$  fraction test to be extended
to $z\simeq 8$.  

\section*{Acknowledgments}
DPS acknowledges financial support from a postdoctoral fellowship from
the Science Technology and Research Council and a Schlumberger
Interdisciplinary  Research Fellow at Darwin College. RSE acknowledges
financial support from the Royal Society.   We acknowledge useful
conversations with George Becker, Jim Dunlop,  Anna Quider, Ross
McLure, Johan Richard, Elizabeth Stanway, and Stephen  Wilkins, and we
thank Sebastiano  Cantalupo for comments after reading an earlier
version of the draft.

\bibliographystyle{mn2e}
\bibliography{mybib}

\begin{thebibliography}{103}
\expandafter\ifx\csname natexlab\endcsname\relax\def\natexlab#1{#1}\fi

\bibitem[{{Ando} {et~al.}(2006){Ando}, {Ohta}, {Iwata}, {Akiyama}, {Aoki}, \&
  {Tamura}}]{Ando06}
{Ando} M., {Ohta} K., {Iwata} I., {Akiyama} M., {Aoki} K., {Tamura} N., 2006,
  \apjl, 645, L9

\bibitem[{{Ando} {et~al.}(2007){Ando}, {Ohta}, {Iwata}, {Akiyama}, {Aoki}, \&
  {Tamura}}]{Ando07}
---, 2007, \pasj, 59, 717

\bibitem[{{Atek} {et~al.}(2008){Atek}, {Kunth}, {Hayes}, {{\"O}stlin}, \&
  {Mas-Hesse}}]{Atek08}
{Atek} H., {Kunth} D., {Hayes} M., {{\"O}stlin} G., {Mas-Hesse} J.~M., 2008,
  \aap, 488, 491

\bibitem[{{Balestra} {et~al.}(2010){Balestra}, {Mainieri}, {Popesso},
  {Dickinson}, {Nonino}, {Rosati}, {Teimoorinia}, {Vanzella}, {Cristiani},
  {Cesarsky}, {Fosbury}, {Kuntschner}, {Rettura}, \& {the GOODS
  team}}]{Balestra10}
{Balestra} I., {Mainieri} V., {Popesso} P., {Dickinson} M., {Nonino} M.,
  {Rosati} P., {Teimoorinia} H., {Vanzella} E., {Cristiani} S., {Cesarsky} C.,
  {Fosbury} R.~A.~E., {Kuntschner} H., {Rettura} A., {the GOODS team}, 2010,
  Accepted for publication in A\&A, arXiV:1001.1115

\bibitem[{{Becker} {et~al.}(2007){Becker}, {Rauch}, \& {Sargent}}]{Becker07}
{Becker} G.~D., {Rauch} M., {Sargent} W.~L.~W., 2007, \apj, 662, 72

\bibitem[{{Bouwens} {et~al.}(2006){Bouwens}, {Illingworth}, {Blakeslee}, \&
  {Franx}}]{Bouwens06a}
{Bouwens} R.~J., {Illingworth} G.~D., {Blakeslee} J.~P., {Franx} M., 2006,
  \apj, 653, 53

\bibitem[{{Bouwens} {et~al.}(2009{\natexlab{a}}){Bouwens}, {Illingworth},
  {Bradley}, {Ford}, {Franx}, {Zheng}, {Broadhurst}, {Coe}, \&
  {Jee}}]{Bouwens09a}
{Bouwens} R.~J., {Illingworth} G.~D., {Bradley} L.~D., {Ford} H., {Franx} M.,
  {Zheng} W., {Broadhurst} T., {Coe} D., {Jee} M.~J., 2009{\natexlab{a}}, \apj,
  690, 1764

\bibitem[{{Bouwens} {et~al.}(2009{\natexlab{b}}){Bouwens}, {Illingworth},
  {Franx}, {Chary}, {Meurer}, {Conselice}, {Ford}, {Giavalisco}, \& {van
  Dokkum}}]{Bouwens09b}
{Bouwens} R.~J., {Illingworth} G.~D., {Franx} M., {Chary} R., {Meurer} G.~R.,
  {Conselice} C.~J., {Ford} H., {Giavalisco} M., {van Dokkum} P.,
  2009{\natexlab{b}}, \apj, 705, 936

\bibitem[{{Bouwens} {et~al.}(2007){Bouwens}, {Illingworth}, {Franx}, \&
  {Ford}}]{Bouwens07}
{Bouwens} R.~J., {Illingworth} G.~D., {Franx} M., {Ford} H., 2007, \apj, 670,
  928

\bibitem[{{Bouwens} {et~al.}(2010{\natexlab{a}}){Bouwens}, {Illingworth},
  {Oesch}, {Stiavelli}, {van Dokkum}, {Trenti}, {Magee}, {Labb{\'e}}, {Franx},
  {Carollo}, \& {Gonzalez}}]{Bouwens10b}
{Bouwens} R.~J., {Illingworth} G.~D., {Oesch} P.~A., {Stiavelli} M., {van
  Dokkum} P., {Trenti} M., {Magee} D., {Labb{\'e}} I., {Franx} M., {Carollo}
  C.~M., {Gonzalez} V., 2010{\natexlab{a}}, \apjl, 709, L133

\bibitem[{{Bouwens} {et~al.}(2010{\natexlab{b}}){Bouwens}, {Illingworth},
  {Oesch}, {Trenti}, {Stiavelli}, {Carollo}, {Franx}, {van Dokkum},
  {Labb{\'e}}, \& {Magee}}]{Bouwens10a}
{Bouwens} R.~J., {Illingworth} G.~D., {Oesch} P.~A., {Trenti} M., {Stiavelli}
  M., {Carollo} C.~M., {Franx} M., {van Dokkum} P.~G., {Labb{\'e}} I., {Magee}
  D., 2010{\natexlab{b}}, \apjl, 708, L69

\bibitem[{{Bunker} {et~al.}(2010){Bunker}, {Wilkins}, {Ellis}, {Stark},
  {Lorenzoni}, {Chiu}, {Lacy}, {Jarvis}, \& {Hickey}}]{Bunker10}
{Bunker} A., {Wilkins} S., {Ellis} R., {Stark} D., {Lorenzoni} S., {Chiu} K.,
  {Lacy} M., {Jarvis} M., {Hickey} S., 2010, Submitted to MNRAS,
  arXiv:0909.2255

\bibitem[{{Bunker} {et~al.}(2004){Bunker}, {Stanway}, {Ellis}, \&
  {McMahon}}]{Bunker04}
{Bunker} A.~J., {Stanway} E.~R., {Ellis} R.~S., {McMahon} R.~G., 2004, \mnras,
  355, 374

\bibitem[{{Bunker} {et~al.}(2003){Bunker}, {Stanway}, {Ellis}, {McMahon}, \&
  {McCarthy}}]{Bunker03}
{Bunker} A.~J., {Stanway} E.~R., {Ellis} R.~S., {McMahon} R.~G., {McCarthy}
  P.~J., 2003, \mnras, 342, L47

\bibitem[{{Charlot} \& {Fall}(1993)}]{Charlot93}
{Charlot} S., {Fall} S.~M., 1993, \apj, 415, 580

\bibitem[{{Coe} {et~al.}(2006){Coe}, {Ben{\'{\i}}tez}, {S{\'a}nchez}, {Jee},
  {Bouwens}, \& {Ford}}]{Coe06}
{Coe} D., {Ben{\'{\i}}tez} N., {S{\'a}nchez} S.~F., {Jee} M., {Bouwens} R.,
  {Ford} H., 2006, \aj, 132, 926

\bibitem[{{Dav{\'e}}(2008)}]{Dave08}
{Dav{\'e}} R., 2008, \mnras, 385, 147

\bibitem[{{Dayal} {et~al.}(2010){Dayal}, {Maselli}, \& {Ferrara}}]{Dayal10}
{Dayal} P., {Maselli} A., {Ferrara} A., 2010, Submitted to MNRAS,
  arXiv/1002.0839

\bibitem[{{Dijkstra} {et~al.}(2007{\natexlab{a}}){Dijkstra}, {Lidz}, \&
  {Wyithe}}]{Dijkstra07a}
{Dijkstra} M., {Lidz} A., {Wyithe} J.~S.~B., 2007{\natexlab{a}}, \mnras, 377,
  1175

\bibitem[{{Dijkstra} {et~al.}(2007{\natexlab{b}}){Dijkstra}, {Wyithe}, \&
  {Haiman}}]{Dijkstra07b}
{Dijkstra} M., {Wyithe} J.~S.~B., {Haiman} Z., 2007{\natexlab{b}}, \mnras, 379,
  253

\bibitem[{{Dow-Hygelund} {et~al.}(2007){Dow-Hygelund}, {Holden}, {Bouwens},
  {Illingworth}, {van der Wel}, {Franx}, {van Dokkum}, {Ford}, {Rosati},
  {Magee}, \& {Zirm}}]{Dow-Hygelund07}
{Dow-Hygelund} C.~C., {Holden} B.~P., {Bouwens} R.~J., {Illingworth} G.~D.,
  {van der Wel} A., {Franx} M., {van Dokkum} P.~G., {Ford} H., {Rosati} P.,
  {Magee} D., {Zirm} A., 2007, \apj, 660, 47

\bibitem[{{Dunkley} {et~al.}(2009){Dunkley}, {Komatsu}, {Nolta}, {Spergel},
  {Larson}, {Hinshaw}, {Page}, {Bennett}, {Gold}, {Jarosik}, {Weiland},
  {Halpern}, {Hill}, {Kogut}, {Limon}, {Meyer}, {Tucker}, {Wollack}, \&
  {Wright}}]{Dunkley09}
{Dunkley} J., {Komatsu} E., {Nolta} M.~R., {Spergel} D.~N., {Larson} D.,
  {Hinshaw} G., {Page} L., {Bennett} C.~L., {Gold} B., {Jarosik} N., {Weiland}
  J.~L., {Halpern} M., {Hill} R.~S., {Kogut} A., {Limon} M., {Meyer} S.~S.,
  {Tucker} G.~S., {Wollack} E., {Wright} E.~L., 2009, \apjs, 180, 306

\bibitem[{{Egami} {et~al.}(2005){Egami}, {Kneib}, {Rieke}, {Ellis}, {Richard},
  {Rigby}, {Papovich}, {Stark}, {Santos}, {Huang}, {Dole}, {Le Floc'h}, \&
  {P{\'e}rez-Gonz{\'a}lez}}]{Egami05}
{Egami} E., {Kneib} J.-P., {Rieke} G.~H., {Ellis} R.~S., {Richard} J., {Rigby}
  J., {Papovich} C., {Stark} D., {Santos} M.~R., {Huang} J.-S., {Dole} H., {Le
  Floc'h} E., {P{\'e}rez-Gonz{\'a}lez} P.~G., 2005, \apjl, 618, L5

\bibitem[{{Eyles} {et~al.}(2007){Eyles}, {Bunker}, {Ellis}, {Lacy}, {Stanway},
  {Stark}, \& {Chiu}}]{Eyles07}
{Eyles} L.~P., {Bunker} A.~J., {Ellis} R.~S., {Lacy} M., {Stanway} E.~R.,
  {Stark} D.~P., {Chiu} K., 2007, \mnras, 374, 910

\bibitem[{{Eyles} {et~al.}(2005){Eyles}, {Bunker}, {Stanway}, {Lacy}, {Ellis},
  \& {Doherty}}]{Eyles05}
{Eyles} L.~P., {Bunker} A.~J., {Stanway} E.~R., {Lacy} M., {Ellis} R.~S.,
  {Doherty} M., 2005, \mnras, 364, 443

\bibitem[{{Faber} {et~al.}(2003){Faber}, {Phillips}, {Kibrick}, {Alcott},
  {Allen}, {Burrous}, {Cantrall}, {Clarke}, {Coil}, {Cowley}, {Davis}, {Deich},
  {Dietsch}, {Gilmore}, {Harper}, {Hilyard}, {Lewis}, {McVeigh}, {Newman},
  {Osborne}, {Schiavon}, {Stover}, {Tucker}, {Wallace}, {Wei}, {Wirth}, \&
  {Wright}}]{Faber03}
{Faber} S.~M., {Phillips} A.~C., {Kibrick} R.~I., {Alcott} B., {Allen} S.~L.,
  {Burrous} J., {Cantrall} T., {Clarke} D., {Coil} A.~L., {Cowley} D.~J.,
  {Davis} M., {Deich} W.~T.~S., {Dietsch} K., {Gilmore} D.~K., {Harper} C.~A.,
  {Hilyard} D.~F., {Lewis} J.~P., {McVeigh} M., {Newman} J., {Osborne} J.,
  {Schiavon} R., {Stover} R.~J., {Tucker} D., {Wallace} V., {Wei} M., {Wirth}
  G., {Wright} C.~A., 2003, in Society of Photo-Optical Instrumentation
  Engineers (SPIE) Conference Series, Vol. 4841, Society of Photo-Optical
  Instrumentation Engineers (SPIE) Conference Series, {M.~Iye \&
  A.~F.~M.~Moorwood}, ed., pp. 1657--1669

\bibitem[{{Fan} {et~al.}(2006){Fan}, {Strauss}, {Becker}, {White}, {Gunn},
  {Knapp}, {Richards}, {Schneider}, {Brinkmann}, \& {Fukugita}}]{Fan06}
{Fan} X., {Strauss} M.~A., {Becker} R.~H., {White} R.~L., {Gunn} J.~E., {Knapp}
  G.~R., {Richards} G.~T., {Schneider} D.~P., {Brinkmann} J., {Fukugita} M.,
  2006, \aj, 132, 117

\bibitem[{{Finkelstein} {et~al.}(2008){Finkelstein}, {Rhoads}, {Malhotra},
  {Grogin}, \& {Wang}}]{Finkelstein08}
{Finkelstein} S.~L., {Rhoads} J.~E., {Malhotra} S., {Grogin} N., {Wang} J.,
  2008, \apj, 678, 655

\bibitem[{{Furlanetto} {et~al.}(2006){Furlanetto}, {Zaldarriaga}, \&
  {Hernquist}}]{Furlanetto06}
{Furlanetto} S.~R., {Zaldarriaga} M., {Hernquist} L., 2006, \mnras, 365, 1012

\bibitem[{{Giavalisco} {et~al.}(2004{\natexlab{a}}){Giavalisco}, {Dickinson},
  {Ferguson}, {Ravindranath}, {Kretchmer}, {Moustakas}, {Madau}, {Fall},
  {Gardner}, {Livio}, {Papovich}, {Renzini}, {Spinrad}, {Stern}, \&
  {Riess}}]{Giavalisco04b}
{Giavalisco} M., {Dickinson} M., {Ferguson} H.~C., {Ravindranath} S.,
  {Kretchmer} C., {Moustakas} L.~A., {Madau} P., {Fall} S.~M., {Gardner} J.~P.,
  {Livio} M., {Papovich} C., {Renzini} A., {Spinrad} H., {Stern} D., {Riess}
  A., 2004{\natexlab{a}}, \apjl, 600, L103

\bibitem[{{Giavalisco} {et~al.}(2004{\natexlab{b}}){Giavalisco}, {Ferguson},
  {Koekemoer}, {Dickinson}, {Alexander}, {Bauer}, {Bergeron}, {Biagetti},
  {Brandt}, {Casertano}, {Cesarsky}, {Chatzichristou}, {Conselice},
  {Cristiani}, {Da Costa}, {Dahlen}, {de Mello}, {Eisenhardt}, {Erben}, {Fall},
  {Fassnacht}, {Fosbury}, {Fruchter}, {Gardner}, {Grogin}, {Hook},
  {Hornschemeier}, {Idzi}, {Jogee}, {Kretchmer}, {Laidler}, {Lee}, {Livio},
  {Lucas}, {Madau}, {Mobasher}, {Moustakas}, {Nonino}, {Padovani}, {Papovich},
  {Park}, {Ravindranath}, {Renzini}, {Richardson}, {Riess}, {Rosati},
  {Schirmer}, {Schreier}, {Somerville}, {Spinrad}, {Stern}, {Stiavelli},
  {Strolger}, {Urry}, {Vandame}, {Williams}, \& {Wolf}}]{Giavalisco04a}
{Giavalisco} M., {Ferguson} H.~C., {Koekemoer} A.~M., {Dickinson} M.,
  {Alexander} D.~M., {Bauer} F.~E., {Bergeron} J., {Biagetti} C., {Brandt}
  W.~N., {Casertano} S., {Cesarsky} C., {Chatzichristou} E., {Conselice} C.,
  {Cristiani} S., {Da Costa} L., {Dahlen} T., {de Mello} D., {Eisenhardt} P.,
  {Erben} T., {Fall} S.~M., {Fassnacht} C., {Fosbury} R., {Fruchter} A.,
  {Gardner} J.~P., {Grogin} N., {Hook} R.~N., {Hornschemeier} A.~E., {Idzi} R.,
  {Jogee} S., {Kretchmer} C., {Laidler} V., {Lee} K.~S., {Livio} M., {Lucas}
  R., {Madau} P., {Mobasher} B., {Moustakas} L.~A., {Nonino} M., {Padovani} P.,
  {Papovich} C., {Park} Y., {Ravindranath} S., {Renzini} A., {Richardson} M.,
  {Riess} A., {Rosati} P., {Schirmer} M., {Schreier} E., {Somerville} R.~S.,
  {Spinrad} H., {Stern} D., {Stiavelli} M., {Strolger} L., {Urry} C.~M.,
  {Vandame} B., {Williams} R., {Wolf} C., 2004{\natexlab{b}}, \apjl, 600, L93

\bibitem[{{Gonz{\'a}lez} {et~al.}(2010){Gonz{\'a}lez}, {Labb{\'e}}, {Bouwens},
  {Illingworth}, {Franx}, {Kriek}, \& {Brammer}}]{Gonzalez09}
{Gonz{\'a}lez} V., {Labb{\'e}} I., {Bouwens} R.~J., {Illingworth} G., {Franx}
  M., {Kriek} M., {Brammer} G.~B., 2010, \apj, 713, 115

\bibitem[{{Haiman} \& {Spaans}(1999)}]{Haiman99}
{Haiman} Z., {Spaans} M., 1999, \apj, 518, 138

\bibitem[{{Hansen} \& {Oh}(2006)}]{Hansen06}
{Hansen} M., {Oh} S.~P., 2006, \mnras, 367, 979

\bibitem[{{Hayes} {et~al.}(2010){Hayes}, {Ostlin}, {Schaerer}, {Mas-Hesse},
  {Leitherer}, {Atek}, {Kunth}, {Verhamme}, {de Barros}, \&
  {Melinder}}]{Hayes10}
{Hayes} M., {Ostlin} G., {Schaerer} D., {Mas-Hesse} J.~M., {Leitherer} C.,
  {Atek} H., {Kunth} D., {Verhamme} A., {de Barros} S., {Melinder} J., 2010,
  ArXiv e-prints

\bibitem[{{Hibon} {et~al.}(2009){Hibon}, {Cuby}, {Willis}, {Cl{\'e}ment},
  {Lidman}, {Arnouts}, {Kneib}, {Willott}, {Marmo}, \& {McCracken}}]{Hibon09}
{Hibon} P., {Cuby} J., {Willis} J., {Cl{\'e}ment} B., {Lidman} C., {Arnouts}
  S., {Kneib} J., {Willott} C.~J., {Marmo} C., {McCracken} H., 2009, Accepted
  in A\&A, arXiv:0907.3354

\bibitem[{{Iliev} {et~al.}(2008){Iliev}, {Shapiro}, {McDonald}, {Mellema}, \&
  {Pen}}]{Iliev08}
{Iliev} I.~T., {Shapiro} P.~R., {McDonald} P., {Mellema} G., {Pen} U., 2008,
  \mnras, 391, 63

\bibitem[{{Iye} {et~al.}(2006){Iye}, {Ota}, {Kashikawa}, {Furusawa},
  {Hashimoto}, {Hattori}, {Matsuda}, {Morokuma}, {Ouchi}, \&
  {Shimasaku}}]{Iye06}
{Iye} M., {Ota} K., {Kashikawa} N., {Furusawa} H., {Hashimoto} T., {Hattori}
  T., {Matsuda} Y., {Morokuma} T., {Ouchi} M., {Shimasaku} K., 2006, \nat, 443,
  186

\bibitem[{{Kashikawa} {et~al.}(2006){Kashikawa}, {Shimasaku}, {Malkan}, {Doi},
  {Matsuda}, {Ouchi}, {Taniguchi}, {Ly}, {Nagao}, {Iye}, {Motohara},
  {Murayama}, {Murozono}, {Nariai}, {Ohta}, {Okamura}, {Sasaki}, {Shioya}, \&
  {Umemura}}]{Kashikawa06}
{Kashikawa} N., {Shimasaku} K., {Malkan} M.~A., {Doi} M., {Matsuda} Y., {Ouchi}
  M., {Taniguchi} Y., {Ly} C., {Nagao} T., {Iye} M., {Motohara} K., {Murayama}
  T., {Murozono} K., {Nariai} K., {Ohta} K., {Okamura} S., {Sasaki} T.,
  {Shioya} Y., {Umemura} M., 2006, \apj, 648, 7

\bibitem[{{Kneib} {et~al.}(2004){Kneib}, {Ellis}, {Santos}, \&
  {Richard}}]{Kneib04}
{Kneib} J.-P., {Ellis} R.~S., {Santos} M.~R., {Richard} J., 2004, \apj, 607,
  697

\bibitem[{{Kornei} {et~al.}(2009){Kornei}, {Shapley}, {Erb}, {Steidel},
  {Reddy}, {Pettini}, \& {Bogosavljevic}}]{Kornei09}
{Kornei} K.~A., {Shapley} A.~E., {Erb} D.~K., {Steidel} C.~C., {Reddy} N.~A.,
  {Pettini} M., {Bogosavljevic} M., 2009, Accepted to ApJ, arXiv:0911.2000

\bibitem[{{Labb{\'e}} {et~al.}(2009){Labb{\'e}}, {Gonzalez}, {Bouwens},
  {Illingworth}, {Franx}, {Trenti}, {Oesch}, {van Dokkum}, {Stiavelli},
  {Carollo}, {Kriek}, \& {Magee}}]{Labbe10}
{Labb{\'e}} I., {Gonzalez} V., {Bouwens} R.~J., {Illingworth} G.~D., {Franx}
  M., {Trenti} M., {Oesch} P.~A., {van Dokkum} P.~G., {Stiavelli} M., {Carollo}
  C.~M., {Kriek} M., {Magee} D., 2009, Submitted to ApJ, arXiv:0911.1356

\bibitem[{{Larson} {et~al.}(2010){Larson}, {Dunkley}, {Hinshaw}, {Komatsu},
  {Nolta}, {Bennett}, {Gold}, {Halpern}, {Hill}, {Jarosik}, {Kogut}, {Limon},
  {Meyer}, {Odegard}, {Page}, {Smith}, {Spergel}, {Tucker}, {Weiland},
  {Wollack}, \& {Wright}}]{Larson10}
{Larson} D., {Dunkley} J., {Hinshaw} G., {Komatsu} E., {Nolta} M.~R., {Bennett}
  C.~L., {Gold} B., {Halpern} M., {Hill} R.~S., {Jarosik} N., {Kogut} A.,
  {Limon} M., {Meyer} S.~S., {Odegard} N., {Page} L., {Smith} K.~M., {Spergel}
  D.~N., {Tucker} G.~S., {Weiland} J.~L., {Wollack} E., {Wright} E.~L., 2010,
  Submitted to ApJ, arXiv:1001.4635

\bibitem[{{Leitherer} {et~al.}(1999){Leitherer}, {Schaerer}, {Goldader},
  {Delgado}, {Robert}, {Kune}, {de Mello}, {Devost}, \&
  {Heckman}}]{Leitherer99}
{Leitherer} C., {Schaerer} D., {Goldader} J.~D., {Delgado} R.~M.~G., {Robert}
  C., {Kune} D.~F., {de Mello} D.~F., {Devost} D., {Heckman} T.~M., 1999,
  \apjs, 123, 3

\bibitem[{{MacArthur} {et~al.}(2008){MacArthur}, {Ellis}, {Treu}, {U}, {Bundy},
  \& {Moran}}]{MacArthur08}
{MacArthur} L.~A., {Ellis} R.~S., {Treu} T., {U} V., {Bundy} K., {Moran} S.,
  2008, \apj, 680, 70

\bibitem[{{Malhotra} \& {Rhoads}(2002)}]{Malhotra02}
{Malhotra} S., {Rhoads} J.~E., 2002, \apjl, 565, L71

\bibitem[{{Malhotra} \& {Rhoads}(2004)}]{Malhotra04}
---, 2004, \apjl, 617, L5

\bibitem[{{McLean} {et~al.}(1998){McLean}, {Becklin}, {Bendiksen}, {Brims},
  {Canfield}, {Figer}, {Graham}, {Hare}, {Lacayanga}, {Larkin}, {Larson},
  {Levenson}, {Magnone}, {Teplitz}, \& {Wong}}]{Mclean98}
{McLean} I.~S., {Becklin} E.~E., {Bendiksen} O., {Brims} G., {Canfield} J.,
  {Figer} D.~F., {Graham} J.~R., {Hare} J., {Lacayanga} F., {Larkin} J.~E.,
  {Larson} S.~B., {Levenson} N., {Magnone} N., {Teplitz} H., {Wong} W., 1998,
  in Proc. SPIE Vol. 3354, p. 566-578, Infrared Astronomical Instrumentation,
  Albert M. Fowler; Ed., {Fowler} A.~M., ed., pp. 566--578

\bibitem[{{McLure} {et~al.}(2010){McLure}, {Dunlop}, {Cirasuolo}, {Koekemoer},
  {Sabbi}, {Stark}, {Targett}, \& {Ellis}}]{McLure10}
{McLure} R.~J., {Dunlop} J.~S., {Cirasuolo} M., {Koekemoer} A.~M., {Sabbi} E.,
  {Stark} D.~P., {Targett} T.~A., {Ellis} R.~S., 2010, \mnras, 126

\bibitem[{{McQuinn} {et~al.}(2007){McQuinn}, {Hernquist}, {Zaldarriaga}, \&
  {Dutta}}]{McQuinn07}
{McQuinn} M., {Hernquist} L., {Zaldarriaga} M., {Dutta} S., 2007, \mnras, 381,
  75

\bibitem[{{Meiksin}(2006)}]{Meiksin06}
{Meiksin} A., 2006, \mnras, 365, 807

\bibitem[{{Mesinger} \& {Furlanetto}(2008)}]{Mesinger08}
{Mesinger} A., {Furlanetto} S.~R., 2008, \mnras, 386, 1990

\bibitem[{{Meurer} {et~al.}(1999){Meurer}, {Heckman}, \& {Calzetti}}]{Meurer99}
{Meurer} G.~R., {Heckman} T.~M., {Calzetti} D., 1999, \apj, 521, 64

\bibitem[{{Neufeld}(1991)}]{Neufeld91}
{Neufeld} D.~A., 1991, \apjl, 370, L85

\bibitem[{{Nilsson} {et~al.}(2009){Nilsson}, {M{\"o}ller-Nilsson}, {M{\o}ller},
  {Fynbo}, \& {Shapley}}]{Nilsson09}
{Nilsson} K.~K., {M{\"o}ller-Nilsson} O., {M{\o}ller} P., {Fynbo} J.~P.~U.,
  {Shapley} A.~E., 2009, \mnras, 400, 232

\bibitem[{{Oesch} {et~al.}(2010){Oesch}, {Bouwens}, {Illingworth}, {Carollo},
  {Franx}, {Labb{\'e}}, {Magee}, {Stiavelli}, {Trenti}, \& {van
  Dokkum}}]{Oesch10}
{Oesch} P.~A., {Bouwens} R.~J., {Illingworth} G.~D., {Carollo} C.~M., {Franx}
  M., {Labb{\'e}} I., {Magee} D., {Stiavelli} M., {Trenti} M., {van Dokkum}
  P.~G., 2010, \apjl, 709, L16

\bibitem[{{Oke} \& {Gunn}(1983)}]{Oke83}
{Oke} J.~B., {Gunn} J.~E., 1983, \apj, 266, 713

\bibitem[{{Ono} {et~al.}(2010){Ono}, {Ouchi}, {Shimasaku}, {Akiyama}, {Dunlop},
  {Farrah}, {Lee}, {McLure}, {Okamura}, \& {Yoshida}}]{Ono10}
{Ono} Y., {Ouchi} M., {Shimasaku} K., {Akiyama} M., {Dunlop} J., {Farrah} D.,
  {Lee} J.~C., {McLure} R., {Okamura} S., {Yoshida} M., 2010, \mnras, 402, 1580

\bibitem[{{Ota} {et~al.}(2008){Ota}, {Iye}, {Kashikawa}, {Shimasaku},
  {Kobayashi}, {Totani}, {Nagashima}, {Morokuma}, {Furusawa}, {Hattori},
  {Matsuda}, {Hashimoto}, \& {Ouchi}}]{Ota08}
{Ota} K., {Iye} M., {Kashikawa} N., {Shimasaku} K., {Kobayashi} M., {Totani}
  T., {Nagashima} M., {Morokuma} T., {Furusawa} H., {Hattori} T., {Matsuda} Y.,
  {Hashimoto} T., {Ouchi} M., 2008, \apj, 677, 12

\bibitem[{{Ouchi} {et~al.}(2009){Ouchi}, {Mobasher}, {Shimasaku}, {Ferguson},
  {Fall}, {Ono}, {Kashikawa}, {Morokuma}, {Nakajima}, {Okamura}, {Dickinson},
  {Giavalisco}, \& {Ohta}}]{Ouchi09}
{Ouchi} M., {Mobasher} B., {Shimasaku} K., {Ferguson} H.~C., {Fall} S.~M.,
  {Ono} Y., {Kashikawa} N., {Morokuma} T., {Nakajima} K., {Okamura} S.,
  {Dickinson} M., {Giavalisco} M., {Ohta} K., 2009, \apj, 706, 1136

\bibitem[{{Ouchi} {et~al.}(2008){Ouchi}, {Shimasaku}, {Akiyama}, {Simpson},
  {Saito}, {Ueda}, {Furusawa}, {Sekiguchi}, {Yamada}, {Kodama}, {Kashikawa},
  {Okamura}, {Iye}, {Takata}, {Yoshida}, \& {Yoshida}}]{Ouchi08}
{Ouchi} M., {Shimasaku} K., {Akiyama} M., {Simpson} C., {Saito} T., {Ueda} Y.,
  {Furusawa} H., {Sekiguchi} K., {Yamada} T., {Kodama} T., {Kashikawa} N.,
  {Okamura} S., {Iye} M., {Takata} T., {Yoshida} M., {Yoshida} M., 2008, \apjs,
  176, 301

\bibitem[{{Ouchi} {et~al.}(2004){Ouchi}, {Shimasaku}, {Okamura}, {Furusawa},
  {Kashikawa}, {Ota}, {Doi}, {Hamabe}, {Kimura}, {Komiyama}, {Miyazaki},
  {Miyazaki}, {Nakata}, {Sekiguchi}, {Yagi}, \& {Yasuda}}]{Ouchi04a}
{Ouchi} M., {Shimasaku} K., {Okamura} S., {Furusawa} H., {Kashikawa} N., {Ota}
  K., {Doi} M., {Hamabe} M., {Kimura} M., {Komiyama} Y., {Miyazaki} M.,
  {Miyazaki} S., {Nakata} F., {Sekiguchi} M., {Yagi} M., {Yasuda} N., 2004,
  \apj, 611, 660

\bibitem[{{Pentericci} {et~al.}(2009){Pentericci}, {Grazian}, {Fontana},
  {Castellano}, {Giallongo}, {Salimbeni}, \& {Santini}}]{Pentericci09}
{Pentericci} L., {Grazian} A., {Fontana} A., {Castellano} M., {Giallongo} E.,
  {Salimbeni} S., {Santini} P., 2009, \aap, 494, 553

\bibitem[{{Pirzkal} {et~al.}(2007){Pirzkal}, {Malhotra}, {Rhoads}, \&
  {Xu}}]{Pirzkal07}
{Pirzkal} N., {Malhotra} S., {Rhoads} J.~E., {Xu} C., 2007, \apj, 667, 49

\bibitem[{{Quider} {et~al.}(2009){Quider}, {Pettini}, {Shapley}, \&
  {Steidel}}]{Quider09}
{Quider} A.~M., {Pettini} M., {Shapley} A.~E., {Steidel} C.~C., 2009, \mnras,
  398, 1263

\bibitem[{{Quider} {et~al.}(2010){Quider}, {Shapley}, {Pettini}, {Steidel}, \&
  {Stark}}]{Quider10}
{Quider} A.~M., {Shapley} A.~E., {Pettini} M., {Steidel} C.~C., {Stark} D.~P.,
  2010, \mnras, 25

\bibitem[{{Reddy} \& {Steidel}(2009)}]{Reddy09}
{Reddy} N.~A., {Steidel} C.~C., 2009, \apj, 692, 778

\bibitem[{{Rhoads} \& {Malhotra}(2001)}]{Rhoads01}
{Rhoads} J.~E., {Malhotra} S., 2001, \apjl, 563, L5

\bibitem[{{Rhoads} {et~al.}(2009){Rhoads}, {Malhotra}, {Pirzkal}, {Dickinson},
  {Cohen}, {Grogin}, {Hathi}, {Xu}, {Ferreras}, {Gronwall}, {Koekemoer},
  {K{\"u}mmel}, {Meurer}, {Panagia}, {Pasquali}, {Ryan}, {Straughn}, {Walsh},
  {Windhorst}, \& {Yan}}]{Rhoads09}
{Rhoads} J.~E., {Malhotra} S., {Pirzkal} N., {Dickinson} M., {Cohen} S.,
  {Grogin} N., {Hathi} N., {Xu} C., {Ferreras} I., {Gronwall} C., {Koekemoer}
  A., {K{\"u}mmel} M., {Meurer} G., {Panagia} N., {Pasquali} A., {Ryan} R.,
  {Straughn} A., {Walsh} J., {Windhorst} R.~A., {Yan} H., 2009, \apj, 697, 942

\bibitem[{{Richard} {et~al.}(2008){Richard}, {Stark}, {Ellis}, {George},
  {Egami}, {Kneib}, \& {Smith}}]{Richard08}
{Richard} J., {Stark} D.~P., {Ellis} R.~S., {George} M.~R., {Egami} E., {Kneib}
  J., {Smith} G.~P., 2008, \apj, 685, 705

\bibitem[{{Salvadori} \& {Ferrara}(2009)}]{Salvadori09}
{Salvadori} S., {Ferrara} A., 2009, \mnras, 395, L6

\bibitem[{{Santos}(2004)}]{Santos04a}
{Santos} M.~R., 2004, \mnras, 349, 1137

\bibitem[{{Schaerer}(2003)}]{Schaerer03}
{Schaerer} D., 2003, \aap, 397, 527

\bibitem[{{Schaerer} \& {de Barros}(2009)}]{Schaerer09}
{Schaerer} D., {de Barros} S., 2009, \aap, 502, 423

\bibitem[{{Schaerer} \& {Verhamme}(2008)}]{Schaerer08}
{Schaerer} D., {Verhamme} A., 2008, \aap, 480, 369

\bibitem[{{Shapley} {et~al.}(2001){Shapley}, {Steidel}, {Adelberger},
  {Dickinson}, {Giavalisco}, \& {Pettini}}]{Shapley01}
{Shapley} A.~E., {Steidel} C.~C., {Adelberger} K.~L., {Dickinson} M.,
  {Giavalisco} M., {Pettini} M., 2001, \apj, 562, 95

\bibitem[{{Shapley} {et~al.}(2003){Shapley}, {Steidel}, {Pettini}, \&
  {Adelberger}}]{Shapley03}
{Shapley} A.~E., {Steidel} C.~C., {Pettini} M., {Adelberger} K.~L., 2003, \apj,
  588, 65

\bibitem[{{Shapley} {et~al.}(2006){Shapley}, {Steidel}, {Pettini},
  {Adelberger}, \& {Erb}}]{Shapley06}
{Shapley} A.~E., {Steidel} C.~C., {Pettini} M., {Adelberger} K.~L., {Erb}
  D.~K., 2006, Accepted for Publication in ApJ

\bibitem[{{Shimasaku} {et~al.}(2006){Shimasaku}, {Kashikawa}, {Doi}, {Ly},
  {Malkan}, {Matsuda}, {Ouchi}, {Hayashino}, {Iye}, {Motohara}, {Murayama},
  {Nagao}, {Ohta}, {Okamura}, {Sasaki}, {Shioya}, \& {Taniguchi}}]{Shimasaku06}
{Shimasaku} K., {Kashikawa} N., {Doi} M., {Ly} C., {Malkan} M.~A., {Matsuda}
  Y., {Ouchi} M., {Hayashino} T., {Iye} M., {Motohara} K., {Murayama} T.,
  {Nagao} T., {Ohta} K., {Okamura} S., {Sasaki} T., {Shioya} Y., {Taniguchi}
  Y., 2006, \pasj, 58, 313

\bibitem[{{Stanway} {et~al.}(2008){Stanway}, {Bremer}, \&
  {Lehnert}}]{Stanway08}
{Stanway} E.~R., {Bremer} M.~N., {Lehnert} M.~D., 2008, \mnras, 385, 493

\bibitem[{{Stanway} {et~al.}(2007){Stanway}, {Bunker}, {Glazebrook}, {Abraham},
  {Rhoads}, {Malhotra}, {Crampton}, {Colless}, \& {Chiu}}]{Stanway07}
{Stanway} E.~R., {Bunker} A.~J., {Glazebrook} K., {Abraham} R.~G., {Rhoads} J.,
  {Malhotra} S., {Crampton} D., {Colless} M., {Chiu} K., 2007, \mnras, 376, 727

\bibitem[{{Stanway} {et~al.}(2003){Stanway}, {Bunker}, \&
  {McMahon}}]{Stanway03}
{Stanway} E.~R., {Bunker} A.~J., {McMahon} R.~G., 2003, \mnras, 342, 439

\bibitem[{{Stanway} {et~al.}(2004){Stanway}, {Bunker}, {McMahon}, {Ellis},
  {Treu}, \& {McCarthy}}]{Stanway04}
{Stanway} E.~R., {Bunker} A.~J., {McMahon} R.~G., {Ellis} R.~S., {Treu} T.,
  {McCarthy} P.~J., 2004, \apj, 607, 704

\bibitem[{{Stanway} {et~al.}(2005){Stanway}, {McMahon}, \&
  {Bunker}}]{Stanway05}
{Stanway} E.~R., {McMahon} R.~G., {Bunker} A.~J., 2005, \mnras, 359, 1184

\bibitem[{{Stark} {et~al.}(2007{\natexlab{a}}){Stark}, {Bunker}, {Ellis},
  {Eyles}, \& {Lacy}}]{Stark07a}
{Stark} D.~P., {Bunker} A.~J., {Ellis} R.~S., {Eyles} L.~P., {Lacy} M.,
  2007{\natexlab{a}}, \apj, 659, 84

\bibitem[{{Stark} {et~al.}(2009){Stark}, {Ellis}, {Bunker}, {Bundy}, {Targett},
  {Benson}, \& {Lacy}}]{Stark09}
{Stark} D.~P., {Ellis} R.~S., {Bunker} A., {Bundy} K., {Targett} T., {Benson}
  A., {Lacy} M., 2009, \apj, 697, 1493

\bibitem[{{Stark} {et~al.}(2007{\natexlab{b}}){Stark}, {Ellis}, {Richard},
  {Kneib}, {Smith}, \& {Santos}}]{Stark07b}
{Stark} D.~P., {Ellis} R.~S., {Richard} J., {Kneib} J., {Smith} G.~P., {Santos}
  M.~R., 2007{\natexlab{b}}, \apj, 663, 10

\bibitem[{{Steidel} {et~al.}(1999){Steidel}, {Adelberger}, {Giavalisco},
  {Dickinson}, \& {Pettini}}]{Steidel99}
{Steidel} C.~C., {Adelberger} K.~L., {Giavalisco} M., {Dickinson} M., {Pettini}
  M., 1999, \apj, 519, 1

\bibitem[{{Steidel} {et~al.}(2000){Steidel}, {Adelberger}, {Shapley},
  {Pettini}, {Dickinson}, \& {Giavalisco}}]{Steidel00}
{Steidel} C.~C., {Adelberger} K.~L., {Shapley} A.~E., {Pettini} M., {Dickinson}
  M., {Giavalisco} M., 2000, \apj, 532, 170

\bibitem[{{Steidel} {et~al.}(2003){Steidel}, {Adelberger}, {Shapley},
  {Pettini}, {Dickinson}, \& {Giavalisco}}]{Steidel03}
---, 2003, \apj, 592, 728

\bibitem[{{Steidel} {et~al.}(2010){Steidel}, {Erb}, {Shapley}, {Pettini},
  {Reddy}, {Bogosavljevi{\'c}}, {Rudie}, \& {Rakic}}]{Steidel10}
{Steidel} C.~C., {Erb} D.~K., {Shapley} A.~E., {Pettini} M., {Reddy} N.~A.,
  {Bogosavljevi{\'c}} M., {Rudie} G.~C., {Rakic} O., 2010, Submitted to ApJ,
  arXiV:/1003.0679

\bibitem[{{Vanzella} {et~al.}(2002){Vanzella}, {Cristiani}, {Arnouts},
  {Dennefeld}, {Fontana}, {Grazian}, {Nonino}, {Petitjean}, \&
  {Saracco}}]{Vanzella02}
{Vanzella} E., {Cristiani} S., {Arnouts} S., {Dennefeld} M., {Fontana} A.,
  {Grazian} A., {Nonino} M., {Petitjean} P., {Saracco} P., 2002, \aap, 396, 847

\bibitem[{{Vanzella} {et~al.}(2008){Vanzella}, {Cristiani}, {Dickinson},
  {Giavalisco}, {Kuntschner}, {Haase}, {Nonino}, {Rosati}, {Cesarsky},
  {Ferguson}, {Fosbury}, {Grazian}, {Moustakas}, {Rettura}, {Popesso},
  {Renzini}, {Stern}, \& {GOODS Team}}]{Vanzella08}
{Vanzella} E., {Cristiani} S., {Dickinson} M., {Giavalisco} M., {Kuntschner}
  H., {Haase} J., {Nonino} M., {Rosati} P., {Cesarsky} C., {Ferguson} H.~C.,
  {Fosbury} R.~A.~E., {Grazian} A., {Moustakas} L.~A., {Rettura} A., {Popesso}
  P., {Renzini} A., {Stern} D., {GOODS Team}, 2008, \aap, 478, 83

\bibitem[{{Vanzella} {et~al.}(2005){Vanzella}, {Cristiani}, {Dickinson},
  {Kuntschner}, {Moustakas}, {Nonino}, {Rosati}, {Stern}, {Cesarsky}, {Ettori},
  {Ferguson}, {Fosbury}, {Giavalisco}, {Haase}, {Renzini}, {Rettura}, {Serra},
  \& {The Goods Team}}]{Vanzella05}
{Vanzella} E., {Cristiani} S., {Dickinson} M., {Kuntschner} H., {Moustakas}
  L.~A., {Nonino} M., {Rosati} P., {Stern} D., {Cesarsky} C., {Ettori} S.,
  {Ferguson} H.~C., {Fosbury} R.~A.~E., {Giavalisco} M., {Haase} J., {Renzini}
  A., {Rettura} A., {Serra} P., {The Goods Team}, 2005, \aap, 434, 53

\bibitem[{{Vanzella} {et~al.}(2006){Vanzella}, {Cristiani}, {Dickinson},
  {Kuntschner}, {Nonino}, {Rettura}, {Rosati}, {Vernet}, {Cesarsky},
  {Ferguson}, {Fosbury}, {Giavalisco}, {Grazian}, {Haase}, {Moustakas},
  {Popesso}, {Renzini}, {Stern}, \& {GOODS Team}}]{Vanzella06}
{Vanzella} E., {Cristiani} S., {Dickinson} M., {Kuntschner} H., {Nonino} M.,
  {Rettura} A., {Rosati} P., {Vernet} J., {Cesarsky} C., {Ferguson} H.~C.,
  {Fosbury} R.~A.~E., {Giavalisco} M., {Grazian} A., {Haase} J., {Moustakas}
  L.~A., {Popesso} P., {Renzini} A., {Stern} D., {GOODS Team}, 2006, \aap, 454,
  423

\bibitem[{{Vanzella} {et~al.}(2009){Vanzella}, {Giavalisco}, {Dickinson},
  {Cristiani}, {Nonino}, {Kuntschner}, {Popesso}, {Rosati}, {Renzini}, {Stern},
  {Cesarsky}, {Ferguson}, \& {Fosbury}}]{Vanzella09}
{Vanzella} E., {Giavalisco} M., {Dickinson} M., {Cristiani} S., {Nonino} M.,
  {Kuntschner} H., {Popesso} P., {Rosati} P., {Renzini} A., {Stern} D.,
  {Cesarsky} C., {Ferguson} H.~C., {Fosbury} R.~A.~E., 2009, \apj, 695, 1163

\bibitem[{{Verhamme} {et~al.}(2008){Verhamme}, {Schaerer}, {Atek}, \&
  {Tapken}}]{Verhamme08}
{Verhamme} A., {Schaerer} D., {Atek} H., {Tapken} C., 2008, \aap, 491, 89

\bibitem[{{Verhamme} {et~al.}(2006){Verhamme}, {Schaerer}, \&
  {Maselli}}]{Verhamme06}
{Verhamme} A., {Schaerer} D., {Maselli} A., 2006, \aap, 460, 397

\bibitem[{{Wilkins} {et~al.}(2010{\natexlab{a}}){Wilkins}, {Bunker}, {Ellis},
  {Stark}, {Stanway}, {Chiu}, {Lorenzoni}, \& {Jarvis}}]{Wilkins10}
{Wilkins} S.~M., {Bunker} A.~J., {Ellis} R.~S., {Stark} D., {Stanway} E.~R.,
  {Chiu} K., {Lorenzoni} S., {Jarvis} M.~J., 2010{\natexlab{a}}, Accepted for
  publication in MNRAS, arXiv:0910.1098

\bibitem[{{Wilkins} {et~al.}(2010{\natexlab{b}}){Wilkins}, {Bunker},
  {Lorenzoni}, \& {Caruana}}]{Wilkins10b}
{Wilkins} S.~M., {Bunker} A.~J., {Lorenzoni} S., {Caruana} J.,
  2010{\natexlab{b}}, Submitted to MNRAS, arXiv:1002.4866

\bibitem[{{Yan} {et~al.}(2006){Yan}, {Dickinson}, {Giavalisco}, {Stern},
  {Eisenhardt}, \& {Ferguson}}]{HYan06}
{Yan} H., {Dickinson} M., {Giavalisco} M., {Stern} D., {Eisenhardt} P.~R.~M.,
  {Ferguson} H.~C., 2006, ArXiv Astrophysics e-prints

\bibitem[{{Yoshida} {et~al.}(2006){Yoshida}, {Shimasaku}, {Kashikawa}, {Ouchi},
  {Okamura}, {Ajiki}, {Akiyama}, {Ando}, {Aoki}, {Doi}, {Furusawa},
  {Hayashino}, {Iwamuro}, {Iye}, {Karoji}, {Kobayashi}, {Kodaira}, {Kodama},
  {Komiyama}, {Malkan}, {Matsuda}, {Miyazaki}, {Mizumoto}, {Morokuma},
  {Motohara}, {Murayama}, {Nagao}, {Nariai}, {Ohta}, {Sasaki}, {Sato},
  {Sekiguchi}, {Shioya}, {Tamura}, {Taniguchi}, {Umemura}, {Yamada}, \&
  {Yasuda}}]{Yoshida06}
{Yoshida} M., {Shimasaku} K., {Kashikawa} N., {Ouchi} M., {Okamura} S., {Ajiki}
  M., {Akiyama} M., {Ando} H., {Aoki} K., {Doi} M., {Furusawa} H., {Hayashino}
  T., {Iwamuro} F., {Iye} M., {Karoji} H., {Kobayashi} N., {Kodaira} K.,
  {Kodama} T., {Komiyama} Y., {Malkan} M.~A., {Matsuda} Y., {Miyazaki} S.,
  {Mizumoto} Y., {Morokuma} T., {Motohara} K., {Murayama} T., {Nagao} T.,
  {Nariai} K., {Ohta} K., {Sasaki} T., {Sato} Y., {Sekiguchi} K., {Shioya} Y.,
  {Tamura} H., {Taniguchi} Y., {Umemura} M., {Yamada} T., {Yasuda} N., 2006,
  \apj, 653, 988

\bibitem[{{Zheng} {et~al.}(2009){Zheng}, {Cen}, {Trac}, \&
  {Miralda-Escude}}]{Zheng10}
{Zheng} Z., {Cen} R., {Trac} H., {Miralda-Escude} J., 2009, Submitted to ApJ,
  arXiV:0910.2712

\end{thebibliography}

\label{lastpage}

\end{document}